\documentclass[peerreview,onecolumn,11pt]{IEEEtran}
\usepackage{comment}
\usepackage{color}
\usepackage[table]{xcolor}
\usepackage{subfigure}
\usepackage{amsmath}
\usepackage{graphicx}
\usepackage{verbatim}
\usepackage[ansinew]{inputenc}
\usepackage{epsfig}
\usepackage{amsfonts}
\usepackage{amssymb}
\usepackage{setspace}
\usepackage{soul}
\usepackage{multirow}
\usepackage{wrapfig}

\usepackage{cite} 
\usepackage{url}  

\usepackage{ifthen}  
\usepackage{multicol}   
\usepackage[]{inputenc} 

\usepackage[rotateright]{rotating}
\usepackage[hyperfootnotes=false]{hyperref}
\hypersetup{
  colorlinks   = true, 
  urlcolor     = blue, 
  linkcolor    = black, 
  citecolor   = blue 
}
\ifCLASSINFOpdf
\else
\fi
\renewcommand{\baselinestretch}{1.3}

\begin{document}
%
\title{MU-MIMO MAC Protocols for Wireless Local \\Area Networks: A Survey}

\author{Ruizhi~Liao,
        Boris~Bellalta,~\IEEEmembership{Senior Member,~IEEE,}
        Miquel~Oliver,~\IEEEmembership{Senior Member,~IEEE,}
        and~Zhisheng~Niu,~\IEEEmembership{Fellow,~IEEE}
\thanks{R. Liao, B. Bellalta and M. Oliver are with NeTS Research Group, Department of Information and Communication Technologies, Universitat Pompeu Fabra, Barcelona, 08018, Spain
 (e-mail: \{ruizhi.liao, boris.bellalta, miquel.oliver\}@upf.edu).}
\thanks{Z. Niu is with NiuLab, Department Of Electronic Engineering, Tsinghua Unisersity, Beijing, 100084, China
 (e-mail: niuzhs@tsinghua.edu.cn).}
}

%

\maketitle

%

\begin{abstract}

As wireless devices boom, and bandwidth-hungry applications (e.g., video and cloud uploading) get popular, today's Wireless Local Area Networks (WLANs) become not only crowded but also stressed at throughput. Multi-user Multiple-Input and Multiple-Output (MU-MIMO), an advanced form of MIMO, has gained attention due to its huge potential in improving the performance of WLANs. 

This paper surveys random access based MAC protocols for MU-MIMO enabled WLANs. It first provides background information about the evolution and the fundamental MAC schemes of IEEE 802.11 Standards and Amendments, and then identifies the key requirements of designing MU-MIMO MAC protocols for WLANs. After that, the most representative MU-MIMO MAC proposals in the literature are overviewed by benchmarking their MAC procedures and examining the key components, such as the channel state information acquisition, de/pre-coding and scheduling schemes. Classifications and discussions on important findings of the surveyed MAC protocols are provided, based on which, the research challenges for designing effective MU-MIMO MAC protocols, as well as the envisaged MAC's role in the future heterogeneous networks, are highlighted.

\end{abstract}


\IEEEpeerreviewmaketitle

%

\section{Introduction}\label{sec:intro}

\IEEEPARstart{I}{EEE} 802.11 is a set of Physical Layer (PHY) and Medium Access Control (MAC) specifications for the prevalent Wireless Local Area Networks (WLANs). Current IEEE 802.11 WLANs contribute to approximate $40 \%$ of overall Internet traffic \cite{ciscoTrend}. As wireless devices rapidly increase, and the wireless transmission evolves towards gigabits per second, IEEE 802.11 WLANs are set to dominate the way of Internet access at homes and working places in the future.


Multi-user Multiple-Input and Multiple-Output (MU-MIMO), introduced by IEEE 802.11ac \cite{IEEEac}, is one of the most crucial techniques that lead WLANs towards the gigabit era. Compared to Single-user MIMO (SU-MIMO), which focuses on transmitting to a single destination, MU-MIMO holds the following three advantages: (1) The increased throughput. By employing SU-MIMO, the theoretical capacity gain can be manifested by a multiplicative factor of $min\{N_\text{t}, N_\text{r}\}$, where $N_\text{t}$ and $N_\text{r}$ are the number of transmitting and receiving antennas \cite{goldsmith2003capacity}\cite{gcaicl}; while in the case of MU-MIMO, the multiplicative factor can be further extended to $min\{aN_\text{t}, bN_\text{r}\}$, where $a$ and $b$ are the number of simultaneous transmitters and receivers; (2) The increased diversity gain. The spatially distributed STAs make the MU-MIMO system more immune to the channel rank loss and the antenna correlation \cite{dgess}, which may affect the SU-MIMO system performance by limiting the available transmission rates; (3) The reduced terminal cost. The MU-MIMO system supports multiple spatially separated STAs (even only equipped with a single antenna) to simultaneously communicate with the Access Point (AP), which makes the development of compact and low-cost user terminals possible.

Considerable research efforts have been made to approach the MIMO capacity at the PHY layer \cite{spencer2004zero}\cite{cbpav}, and a comprehensive overview can be found in \cite{jmie}. Following the PHY advance, corresponding MAC enhancements, especially the MU-MIMO based ones, have sprung up. 

MAC, used among multiple stations (STAs) to share a common wireless channel, can be dated back to the pioneer protocols used in the ALOHA network in the $1970$s \cite{abramson1970aloha} to the current Carrier Sense Multiple Access with Collision Avoidance (CSMA/CA) based IEEE 802.11 Distributed Coordination Function (DCF). Since the IEEE 802.11 MAC mechanism only supports one single transmission at a time, which underutilizes the full potential of the spatial domain of MU-MIMO transmissions, thus, MU-MIMO MAC proposals have tried to adapt the frame structure, as well as their operation procedures to control the parallel transmissions among STAs.

There are two main MAC categories: (1) the fixed-assignment one, where the channel frequencies, the access time, mutually orthogonal codes or different polarization is predefined for each STA, namely, Frequency Division, Time Division, Code Division or Polarization Division Multiple Access (FDMA, TDMA, CDMA and PDMA); and (2) the random access one, where each STA independently determines when to compete for the channel, e.g., Aloha and CSMA/CA. It is worth to note that Space Division Multiple Access (SDMA) is a medium access scheme by exploring parallel spatial streams. In this sense, the MU-MIMO transmission, where multiple spatially-separated nodes are involved in parallel transmissions, is a form of SDMA. Thus, SDMA schemes and MU-MIMO transmissions are used interchangeably in this paper.

Due to the following three reasons, the random access based CSMA/CA has dominated the MAC mechanism of WLANs. First, the backward compatibility. The initial WLAN traffic was sporadic, bursty, and asymmetrical between uplink and downlink. Although in the past decade, the traffic load has increased significantly, and the traffic pattern has evolved from mainly web browsing and file transfers to a wide variety of applications, there are still vast amount of legacy STAs based on CSMA/CA. Secondly, coexistence with other networks. Neighbouring WLANs, wireless sensor networks and Bluetooth personal area networks, that are all operating in the Industrial Scientific Medical (ISM) band, present significant interference to each other. CSMA/CA offers a simple but effective solution (i.e., listen before transmitting) to share the unlicensed spectrum among competing networks. Thirdly, the implementation simplicity. The implementation of a random access scheme is simpler compared to others. Since there is no need to use accurate clocks for synchronization, nor to execute complex functions for scheduling. 


The central point of the paper is to study random access based MAC mechanisms for MU-MIMO enabled WLANs. The main contributions of the paper are threefold. (1) We report the IEEE bodies' MAC progress, as well as survey and categorize the most relevant MU-MIMO MAC proposals in the literature. (2) By doing such review, we identify key requirements for designing efficient MU-MIMO MAC protocols, such as the channel state information (CSI) acquisition, de/pre-coding and scheduling schemes. These requirements are used to benchmark MU-MIMO MAC proposals in the literature. (3) We provide highlights and discussions on important findings and research challenges after each subsection of the literature review, and give our thoughts about possible future directions on the MU-MIMO MAC design, such as uplink MU-MIMO, new backoff algorithms and full-duplex transmissions. 

The rest of the paper is organized as follows. First, Section \ref{sec:wlans} briefly overviews the evolution of IEEE 802.11 standards/amendments and their fundamental MAC mechanisms to clarify the MAC development promoted by the IEEE standard body. Next, Section \ref{sec:mac} identifies the key requirements for designing MU-MIMO MAC protocols in WLANs. Then, Section \ref{sec:survey} surveys and classifies the most prominent MU-MIMO MAC protocols in the literature. Afterwards, Section \ref{sec:future} discusses the research challenges and future directions. Finally, Section \ref{sec:conclusions} concludes the paper.

\newcommand{\tabincell}[2]{\begin{tabular}{@{}#1@{}}#2\end{tabular}}
\begin{table*}[t!!!!!!!]
\caption{{ \bfseries Terms and Abbreviations}}
\newsavebox{\tablebox}
\begin{lrbox}{\tablebox}
\renewcommand\arraystretch{2.3}
{\LARGE
\begin{tabular}{|l|l|l|}
\hline
{Access Category (AC) } & {Additive White Gaussian Noise (AWGN) } & {Network Allocation Vector (NAV)  } \\ \hline
Access Point (AP)  &\tabincell{l}{\vspace{-0.4em}Aggregated MAC Protocol Data Unit\\ \vspace{-0.0em}(A-MPDU)} &\tabincell{l}{\vspace{-0.4em}Orthogonal Frequency Division Multiplexing\\ \vspace{-0.0em}(OFDM)}   \\ \hline

Acknowledgement (ACK)  &\tabincell{l}{\vspace{-0.4em}Carrier Sense Multiple Access with Collision \vspace{-0.0em}\\Avoidance (CSMA/CA)}      &Quadrature Amplitude Modulation (QAM)   \\ \hline

Backoff (BO) &Channel State Information (CSI)  &Quality of Service (QoS) \\ \hline

Clear-to-Send (CTS)   &Code Division Multiple Access (CDMA)   &Reduced Inter Frame Space (RIFS)  \\ \hline

Contention Window (CW)   &Direct Sequence Spread Spectrum (DSSS)    &Short Inter Frame Space (SIFS)    \\ \hline

DCF Inter Frame Space (DIFS)   &Distributed Coordination Function (DCF)    &Signal-Interference-Noise Ratio (SINR)     \\ \hline

Dirty Paper Coding (DPC)    &Enhanced Distribution Channel Access EDCA       &Signal-Noise Ratio (SNR)  \\ \hline

\tabincell{l}{\vspace{-0.4em}Explicit Compressed Feedback\\ \vspace{-0.0em}(ECFB)}     &Frequency Division Multiple Access (FDMA)      &\tabincell{l}{\vspace{-0.4em}Single-user Multiple-Input Multiple-Output\\ \vspace{-0.0em}(SU-MIMO)}         \\ \hline

First-in First-out (FIFO)  &Frequency Hopping Spread Spectrum (FHSS)     & Software Defined Radio (SDR)   \\ \hline

High Throughput (HT)   &Minimum Mean Square Error (MMSE)     &Successive Interference Cancellation (SIC)     \\ \hline

Maximum Likelihood (ML)     &Multi-Packet Reception (MPR)        & Time Division Multiple Access (TDMA)   \\ \hline

Medium Access Control (MAC)  &Multi-Packet Transmission (MPT)       & Transmit Opportunity (TXOP)  \\ \hline

Physical Layer (PHY)  & Multi-user Detection (MUD)     &Very High Throughput (VHT)     \\ \hline

Request-to-Send (RTS)  & Multi-user Interference Cancellation (MUIC)     & Wireless Local Area Networks (WLANs)  \\ \hline

Station (STA)  &  \tabincell{l}{\vspace{-0.4em}Mingle-user Multiple-Input Multiple-Output\\ \vspace{-0.0em}(MU-MIMO)}  & Zero Forcing Beamforming (ZFBF)    \\ \hline

\end{tabular}
}
\label{terms}
\end{lrbox}
\resizebox{1\textwidth}{!}{\usebox{\tablebox}}
\end{table*}

\section{The Evolution of IEEE 802.11 and MAC Schemes}\label{sec:wlans}
This section presents an evolutionary overview of IEEE 802.11 standards/amendments, and also introduces how the IEEE 802.11 specified MAC schemes work. This overview does not go through all aspects of IEEE 802.11 standards/amendments, but focuses on the background information that is closely related to the topic of the paper, namely, medium access control. Due to a considerable amount of technical terms and abbreviations have been used, Table \ref{terms} is given to summarize important terms for the reader's convenience.

\subsection{IEEE 802.11 Standards/Amendments}
\subsubsection{Standards}
Loosely speaking, both standards and amendments can be interchangeably used to refer to different variants of IEEE standards or amendments. However, a more strict nomenclature designates standards as documents with mandatory requirements (denoted as IEEE 802.11 followed by the published year, e.g., IEEE 802.11-2012), and amendments as documents that add to, remove from, or alter material in a portion of existing standards \cite{ieesast} (denoted as IEEE 802.11 followed by a non-capitalized letter or letters, e.g., IEEE 802.11n or 802.11ac).

Since 1997, IEEE has released four standards: 802.11-1997, 802.11-1999, 802.11-2007 and 802.11-2012. IEEE 802.11-2012 \cite{IEEE12} is the latest and the only version that is currently in publication. Standards are continuously updated by amendments, e.g., 802.11-2012 is created by integrating ten amendments such as 802.11n and 802.11p with the base standard 802.11-2007, which was replaced since the release of 802.11-2012. In other words, each standard will be superseded by its successor in its entirety.

\subsubsection{Amendments}
In 1999, two amendments were first introduced: (1) IEEE 802.11a operates in the $5$ GHz band using the Orthogonal Frequency Division Multiplexing (OFDM) modulation with a maximum data rate of $54$ Mbps; (2) IEEE 802.11b operates in the $2.4$ GHz band using the Direct Sequence Spread Spectrum (DSSS) modulation with a maximum data rate of $11$ Mbps. Compared to 802.11-1997, 802.11b substantially increases the data rate (from $2$ Mbps to $11$ Mbps) using the same modulation technique and the frequency band, which made 802.11b the then-definitive WLAN technology. In 2003, IEEE 802.11g, a new amendment working in the $2.4$ GHz band was ratified. It extends 802.11b with a maximum data rate of $54$ Mbps. IEEE 802.11n \cite{IEEE09n}, ratified in 2009, operates in either $2.4$ GHz or $5$ GHz band, boosting the data rate to $150$ Mbps ($600$ Mbps by $4$ streams) by utilizing MIMO.

802.11ac \cite{IEEEac} is the latest IEEE amendment approved in 2013. It is operating exclusively in the $5$ GHz band. Driven by the need for higher speed, 802.11ac aims to provide an aggregated multi-station throughput of at least $1$ gigabit per second, namely, Very High Throughput (VHT) WLANs. Compared to 802.11n, this significant improvement is achieved by introducing novel PHY and MAC features, such as wider bandwidths ($80$ and $160$ MHz), a denser modulation scheme ($256$-QAM: Quadrature Amplitude Modulation), a compulsory frame format (A-MPDU: Aggregated MAC Protocol Data Unit), and most importantly, downlink MU-MIMO transmissions (supporting simultaneous transmissions of up to $4$ STAs with the maximum number of $8$ streams).

Although each amendment is revoked as it is merged into the latest standard, the sign of IEEE 802.11a/b/g/n/ac is often employed by vendors to denote the capability and compatibility of their products.

\begin{table*}[t!!!!!!!]
\caption{{ \bfseries Features of Related IEEE 802.11 Standards/Amendments}}
\begin{lrbox}{\tablebox}
\renewcommand\arraystretch{2.3}
{\LARGE
\begin{tabular}{|c|c|c|c|c|c|}
\hline
{\bfseries \LARGE Version} & {\bfseries \LARGE Description} &  {\bfseries \LARGE Frequency} & {\bfseries \LARGE Max. Data Rate} & {\bfseries \LARGE Modulation}\\ \hline
802.11-1997 & WLAN MAC and PHY Specifications &  20 (MHz) @ 2.4 (GHz) & 2 (Mbps) & DSSS, FHSS\\ \hline

802.11-1999 & Part II WLAN MAC and PHY Specifications &  20 @ 2.4  & 2  & DSSS, FHSS\\ \hline

a & Higher Speed PHY Extension &  20 @ 5 & 54  & OFDM\\ \hline

b & Higher Speed PHY Extension &  20 @ 2.4  & 11 & DSSS\\ \hline

g & Further Higher Data Rate Extension  & 20 @ 2.4 & 54  & OFDM, DSSS\\ \hline

802.11-2007 & Standard Maintenance Revision  & -- &  -- & --\\ \hline

n & High Throughput  & 20, 40 @ 2.4, 5 & 150 x 4 & OFDM\\ \hline

802.11-2012 & Accumulated Maintenance Changes & -- &  --  & --\\ \hline

ac & Very High Throughput  & 20, 40, 80, 160 @ 5 & 866.7 x 8  & OFDM\\ \hline

ax & High Efficiency WLAN (approx. 2019)  & Below 6 GHz & - & OFDM\\ 




\hline
\end{tabular}
}
\label{ieeeParameters}
\end{lrbox}
\resizebox{1\textwidth}{!}{\usebox{\tablebox}}
\end{table*}

\subsubsection{Next Amendment-802.11ax}

In March 2014, IEEE created a new task group-802.11ax \cite{11ax}, aiming at delivering High Efficiency WLANs (HEW) in both indoor and outdoor high density scenarios. 802.11ax will operate in frequency bands between $1$ and $6$ GHz. In terms of performance, the focus of the amendment has shifted from improving the aggregated system throughput to the throughput observed by each STA, targeting at at least four-time per-STA throughput increase comparing to the existing standards and amendments. The ongoing amendment is in its early stage, and is estimated to be finished in 2019. Currently, the study group is discussing the 802.11ax usage models focusing on the dense deployment \cite{lcum}, potential technologies such as full-duplex, uplink MU-MIMO and Massive MIMO \cite{gong2014advanced}, and functional requirements such as the system performance, the spectrum efficiency, the operation bands and the backward compatibility \cite{lwpax}.

The key features of the above mentioned IEEE 802.11 standards/amendments are given in Table \ref{ieeeParameters}, where FHSS stands for Frequency Hopping Spread Spectrum.
\subsubsection{New Frequency Bands of IEEE 802.11}
Besides the traditional frequency bands ($2.4$ and $5$ GHz), IEEE 802.11 has extended to support other bands.

IEEE 802.11ad \cite{11ad}, another VHT WLAN amendment, will operate in the $60$ GHz band and focus on multi-gigabits per second data transmissions in a short range point-to-point links (around $10$ meters). A typical application scenario of 802.11ad is the wireless transmission of lightly compressed or uncompressed high-definition videos for home entertainment systems. Due to $60$ GHz band's characteristics of high propagation loss and high attenuation, directional transmissions and receptions are required.


IEEE 802.11ah \cite{11ah} will operate in the sub-GHz band. The main purposes of the amendment are to introduce power saving and STA grouping mechanisms. A typical use case is a smart metering network with many sensor nodes, where high collision probability and hidden nodes are expected. 802.11ah will partition nodes into groups to save power and to reduce the channel contention by assigning the channel to nodes of a given group at a given time \cite{adame2013capacity}\cite{adame2014ieee}.

Due to the specific purpose of each amendment and unique features of the employed frequency, the MU-MIMO MAC schemes designed for WLANs of the traditional bands can not be directly applied to these new amendments. For example, 802.11ah relies on the highly centralized medium access scheme, while 802.11ad has to utilize the beam sweeping technique to detect STAs rather than the omnidirectional carrier sensing adopted by IEEE 802.11 DCF. Therefore, this paper preserves its focus on MAC proposals for the traditional bands, i.e., $2.4$ and $5$ GHz. However, the potential collaborations at the MAC level between protocols of traditional and new bands will be discussed in Section \ref{sec:future}-Future Directions.

\subsection{IEEE 802.11 Medium Access Control}\label{sec:dcf}
Although IEEE 802.11 has specified three MAC mechanisms, namely, DCF, Point Coordination Function (PCF) and Hybrid Coordination Function Controlled Access (HCCA), this paper only focuses on the distributed and random access based MAC schemes, because PCF and HCCA (i.e., the centralized schemes) are neither widely adopted by the industry nor the academia.
\subsubsection{Distributed Coordination Function}
\begin{figure}[h!!!!!!]
\begin{center}
\subfigure[A successful transmission of basic access]{\includegraphics[scale=0.53]{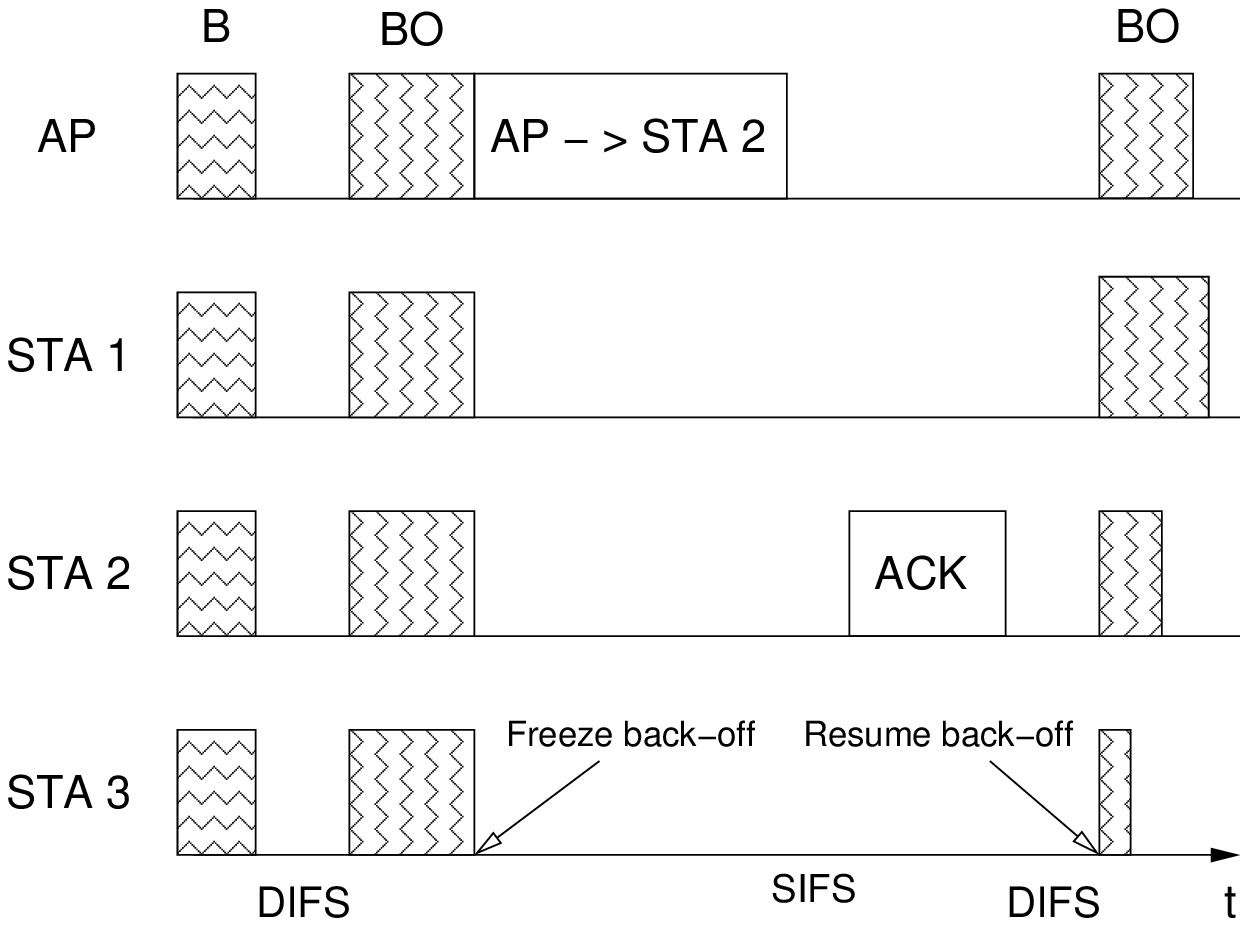}{\label{Fig:Sbas}}}
\subfigure[A successful transmission of RTS/CTS]{\includegraphics[scale=0.53]{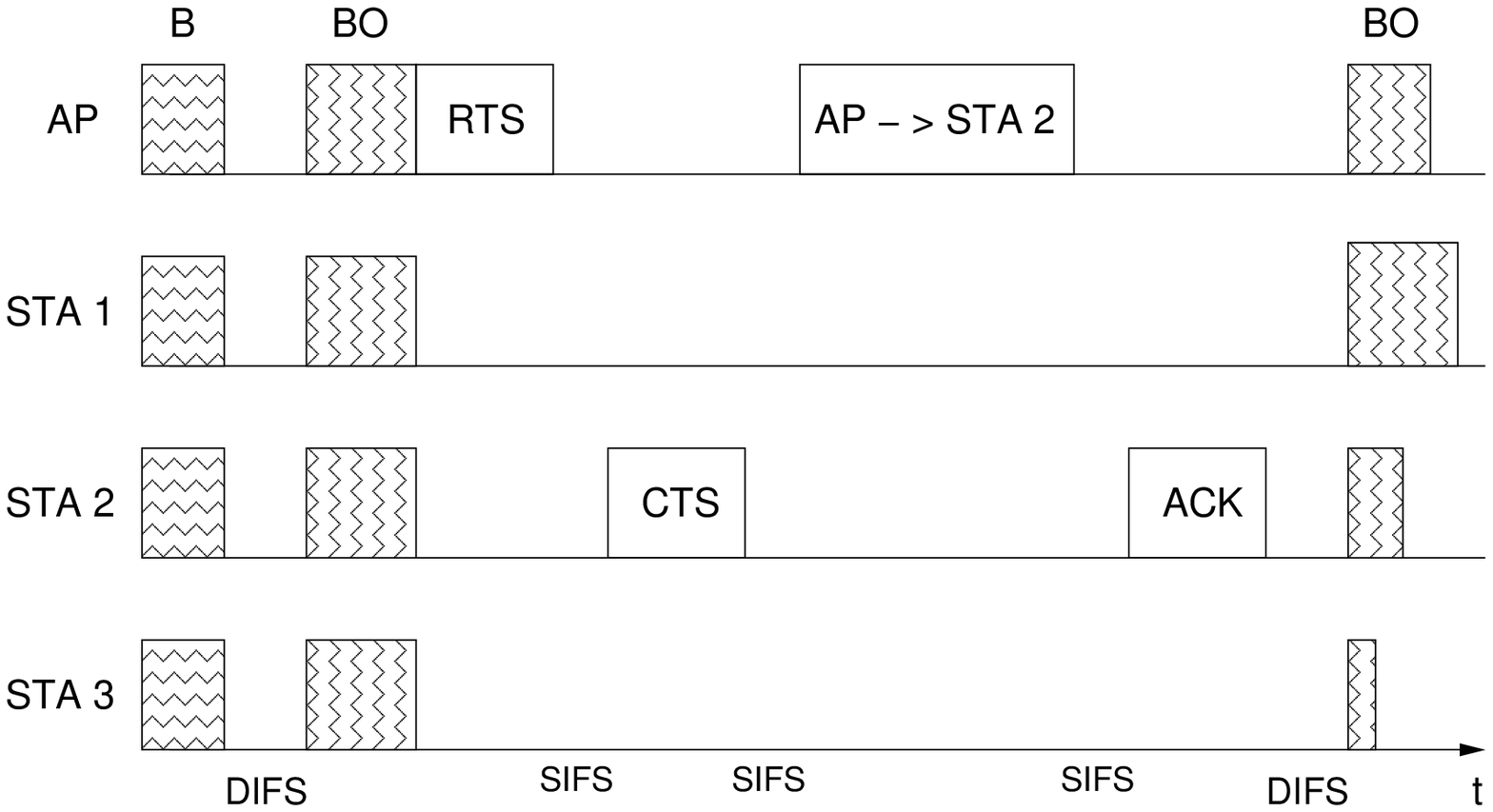}{\label{Fig:Srts}}}
\caption{802.11 DCF transmission procedures}
\label{Fig:diag_S_bas_rts}
\end{center}
\end{figure}

DCF is the fundamental medium access scheme of IEEE 802.11 based WLANs. It relies on CSMA/CA to detect and share the wireless channel among STAs. DCF can either operate in the basic access scheme (Figure \ref{Fig:Sbas}) or the optional Request-to-Send/Clear-to-Send (RTS/CTS, Figure \ref{Fig:Srts}) scheme. DCF mandates STAs to keep sensing the channel. If the channel has been idle for DCF Inter Frame Space (DIFS), each STA starts decreasing a backoff (BO) timer chosen from its Contention Window (CW) to compete for the channel. The STA with the lowest BO wins the channel contention and starts to transmit frames. Collisions occur if more than one STA happens to choose the same random BO. When a transmitted frame is successfully received, the receiver waits for a Short Inter Frame Space (SIFS) and then sends back an Acknowledgement (ACK). Note that as soon as the winning STA sends out a frame, other STAs will notice the channel has become busy, therefore immediately freeze their BO timers. These STAs will wait the channel to be idle for another DIFS, and resume decreasing the remaining BO timers. The STA who previously succeeded the channel contention will have a new BO timer at its next transmission attempt.

Examples of a successful transmission for the basic access and the RTS/CTS schemes are shown in Figure \ref{Fig:diag_S_bas_rts}, where B denotes the channel is initially busy. Please refer to the IEEE standard 802.11-2012 \cite{IEEE12} for more details about DCF. 

\subsubsection{Enhanced Distribution Channel Access}
IEEE 802.11e \cite{IEEE05e} proposes an extension to DCF-Enhanced Distribution Channel Access (EDCA), as a response to the demand of Quality of Service (QoS) for voice and video applications. The main differences between DCF and EDCA are twofold. First, the former does not differentiate traffic from different applications, while the latter classifies traffic into four Access Categories (ACs) with different priorities: Voice (AC\_VO), Video (AC\_VI), Best Effort (AC\_BE) and Background (AC\_BK). By doing so, EDCA is able to assign ACs with different parameters. For example, the maximum Transmit Opportunity (TXOP, a contention-free interval, during which a STA can transmit as many frames as possible) for AC\_VO and AC\_VI are $1.504$ ms and $3.008$ ms, respectively. Secondly, it is also different that the instant of time at which DCF and EDCA mandate STAs to decrease the BO timer. In DCF, STAs decrease the BO timer at the end of each slot, while in EDCA, the decrement occurs at the beginning of each slot. Please refer to \cite{IEEE12} and \cite{gbu8} for detailed comparisons of DCF and EDCA, and \cite{charfi2013phy} for QoS supports in WLANs.

\section{Requirements for Designing MU-MIMO MAC Protocols in WLANs}\label{sec:mac}



MU-MIMO transmissions in WLANs have two communication paths, the uplink one (i.e., STAs simultaneously transmit frames to the AP, which is also referred as the MIMO-MAC channel) and the downlink one (i.e., the AP sends data to a group of STAs in parallel, which is also referred as the MIMO-broadcast channel). The MU-MIMO uplink and downlink transmissions face different challenges, and hence, have different requirements in designing MAC protocols.

\subsection{De/Pre-coding Schemes for Simultaneous Receptions/Transmissions}
\begin{figure}[h!!!!!!!!]
\centering
\includegraphics[scale=0.68]{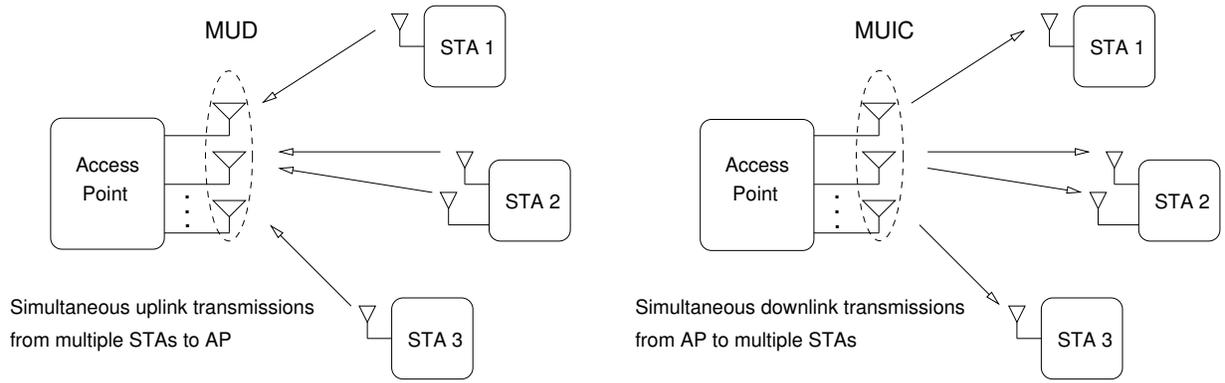}
\caption{Up/Down-link multi-user transmissions}
\label{Fig:updownMUDMUI}
\end{figure}

In the uplink, the AP needs to separate the simultaneously transmitted signals from STAs, which is the Multi-user Detection (MUD) problem. In the downlink, the AP has to, firstly, select a group of STAs based on a certain criterion such as the queue occupancy, given that the selected STAs have to be spatially non-correlated, which is the scheduling problem, and, secondly, precode the outgoing frames to null the interference among concurrent spatial streams, which is the Multi-user Interference Cancellation (MUIC) problem. An illustration of MU-MIMO uplink and downlink transmissions is given in Figure \ref{Fig:updownMUDMUI}.

The design of MUD/MUIC schemes is beyond the topic of the paper. However, some of the most popular MUD/MUIC schemes adopted in the surveyed papers, as well as their strong points and drawbacks, are sampled (Table \ref{mudFeature}).

\subsubsection{MUD Schemes for Simultaneous Uplink Receptions}
\paragraph{Minimum Mean Square Error (MMSE)}
Received signals at each antenna of the AP are multiplied by a complex weight and then summed up. The weight is adjustable through minimizing the difference between the summation of the output signal and a reference that is known by both the AP and STAs. An example of the weight adjustment is to utilize the steepest descending algorithm. The performance of the MMSE MUD scheme improves as the number of AP's antennas increases, and degrades as the network scales up \cite{jiang2007multiuser}.

\paragraph{Maximum Likelihood (ML)}
The ML MUD conducts an exhaustive search to extract the transmitted signals. It provides the best detection performance, but comes with the highest complexity that increases exponentially with the number of STAs, which makes it infeasible in practical systems.

\paragraph{Sphere Decoding (SD)}
Some SD based MUD algorithms have been proposed to reduce the complexity of the pure ML MUD while to approach the performance of ML MUD. The idea is to decrease the radius of the search scope by focusing on the vicinity of the ML solution.

\paragraph{Successive Interference Cancellation (SIC)}
The SIC MUD is an enhancement to MMSE. A detection algorithm is utilized by estimating of the received power at the AP. The signal with the highest power, which is the least interfered by others, is detected. This detected signal is then subtracted from mixed signals, and the next highest signal is singled out using the same process until the lowest STA signal is determined. The SIC MUD tends to be erroneous at the signal classification stage. The false deduction from composite STA signals may propagate the error to following calculations \cite{khalid2010advances}. 

\subsubsection{MUIC Schemes for Simultaneous Downlink Transmissions}
Although simultaneous downlink transmissions from the AP to multiple STAs can be seen as a combination of several single-user transmissions, STAs' random and independent locations make it very challenging to jointly null multi-user interference at the STA side. Therefore, most proposals in the literature precode outgoing signals at the AP to minimize interference among simultaneous streams. 

\paragraph{Zero Forcing (ZF)}
In the ZF scheme, the original signal is multiplied by the pseudo-inverse of the channel matrix to completely null the MUI. The conditions are (1) the AP has the full CSI, and (2) the channel is invertible. By multiplying the pseudo-inverse weight, the ZF scheme also increases the error rate (because the noise vector is amplified). The amplified noise vector indicates that ZF can only perform well in the high Signal-to-Noise Ratio (SNR) region. In addition, the ZF scheme requires that the number of total receiving antennas is not less than that of transmitting antennas \cite{paulraj2003introduction}.

In comparison, the MMSE scheme can minimize the overall error rate without amplifying the noise. \cite{khalid2010advances} and \cite{paulraj2003introduction} show that the MMSE scheme performs better than ZF in the low SNR region, and approaches the performance of ZF in the high SNR region.


\paragraph{Block diagonalization (BD)}
BD is a generalized channel inversion technique, especially when receivers have multiple antennas \cite{spencer2004zero}. Singular value decomposition (SVD) is employed to remove unitary matrices, which makes the computational complexity of BD
higher than MMSE.

\paragraph{Dirty Paper Coding (DPC)}
DPC is a non-linear precoding scheme firstly introduced by Costa \cite{costa1983writing}, which can achieve the optimum performance at the cost of significant computing complexity. The idea is to add an offset (the negative value of the interference that is known at the AP) to the transmitted signal, which hints that (1) the AP has to know the interference in advance, and (2) the AP always has available codewords, i.e., infinite length of codewords, which make DPC not suitable for practical use. 

\begin{table}[h!!!!!!!]
\caption{{\bfseries Features of MUD/MUIC Schemes}}
\begin{lrbox}{\tablebox}
\renewcommand\arraystretch{2.3}
{
\begin{tabular}{|l|l|l|l|}
\hline
{\bfseries Name} & {\bfseries Type} & {\bfseries Main Features}  & {\bfseries Remarks} \\ \hline
Zero Forcing (ZF)&Linear, MUD/MUIC &  \tabincell{l}{\vspace{-0.7em}Complete interference cancellation \vspace{-0.0em}\\with full CSI}  & Noise amplified  \\ \hline

\tabincell{l}{\vspace{-0.7em}Minimum Mean Square\vspace{-0.0em}\\Error (MMSE)} &Linear, MUD/MUIC & \tabincell{l}{\vspace{-0.7em}Complete interference cancellation \vspace{-0.0em}\\with full CSI} & Outperform ZF at low SNR \\ \hline

\tabincell{l}{\vspace{-0.7em}Maximum Likelihood\vspace{-0.0em}\\(ML)}&Nonlinear, MUD & Performance bound of MUD & \tabincell{l}{\vspace{-0.7em}Exhaustive search; \vspace{-0.0em}\\Exponential complexity}   \\ \hline

\tabincell{l}{\vspace{-0.7em}Successive Interference\vspace{-0.0em}\\Cancellation (SIC)} &Nonlinear, MUD & Trade-off between ML \& MMSE   & Error propagation \\ \hline

\tabincell{l}{\vspace{-0.7em}Block diagonalization\vspace{-0.0em}\\(BD)}&Linear, MUIC &More complex than MMSE  & \tabincell{l}{\vspace{-0.7em}SVD; Multi-antenna receivers;\vspace{-0.0em}\\Generalized channel inversion}   \\ \hline

\tabincell{l}{\vspace{-0.7em}Dirty Paper Coding\vspace{-0.0em}\\(DPC)}&Nonlinear, MUIC & Performance bound of MUIC  & \tabincell{l}{\vspace{-0.7em}Infinite codewords;\vspace{-0.0em}\\Impractical for use}  \\ \hline

\end{tabular}
}
\label{mudFeature}
\end{lrbox}
\resizebox{1\textwidth}{!}{\usebox{\tablebox}}
\end{table}

\subsection{Channel State Information Acquisition}
MUD and MUIC schemes allow MU-MIMO systems to separate simultaneously received/transmitted frames, and achieve the spatial multiplexing gain. However, it is important to point out that, in the above discussion, the possession of CSI is assumed at the AP. 

Most proposals in the literature integrate the CSI acquisition into MAC operations. There are generally two types of CSI: the statistical CSI and the instantaneous one. The former employs the statistical characteristics of the channel (e.g., fading distribution, average channel gain and spatial correlation) to decide the CSI, which performs well in scenarios where the channel has a large mean component (i.e., a large Rician factor) or strong correlation (either in space, time, or frequency) \cite{dlova}.

\begin{figure}[h!!!!!!]
\begin{center}
\subfigure[Implicit CSI feedback scheme]{\includegraphics[scale=0.65]{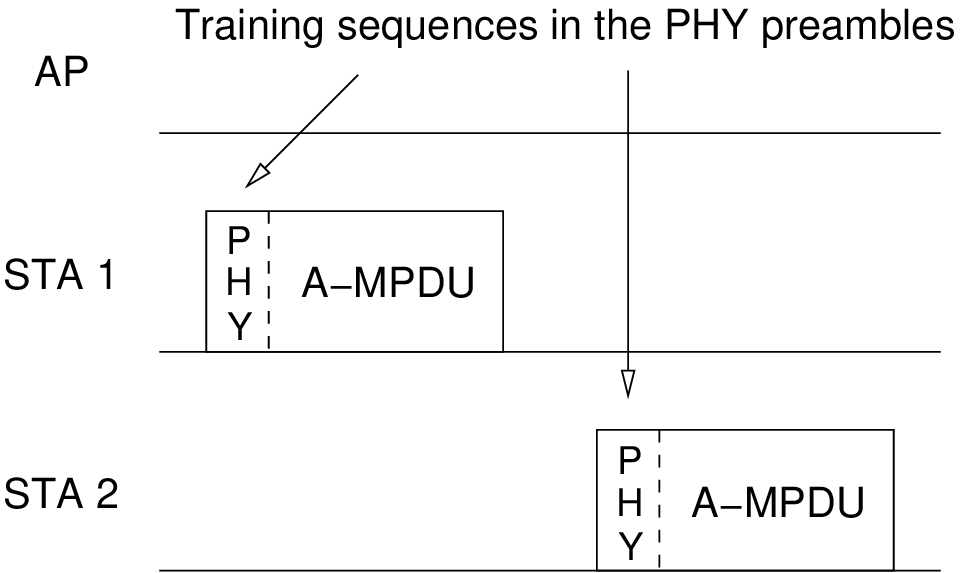}{\label{Fig:imp_csi}}}
\subfigure[Explicit CSI feedback scheme]{\includegraphics[scale=0.65]{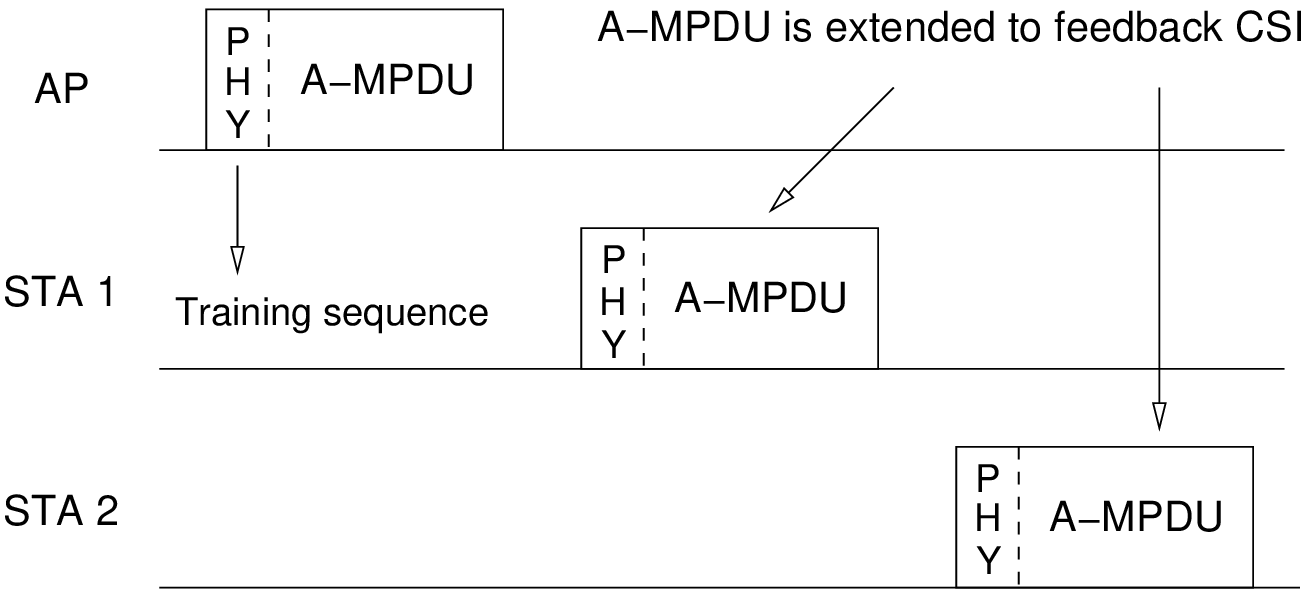}{\label{Fig:exp_csi}}}
\caption{Implicit \& Explicit CSI feedback procedures}
\label{Fig:imp_exp_csi}
\end{center}
\end{figure}

The instantaneous CSI (or the short-term CSI) means the current channel state is known, which enables the transmitter to adapt its outgoing signal. Because wireless channel varies over time, the instantaneous CSI has to be estimated repeatedly on a short-term basis. The acquisition of CSI can be done by estimating a training sequence known by both transmitters and receivers. In the uplink, the AP can easily extract the uplink CSI from the PHY preambles of received frames. While, for transmissions in the downlink, the acquisition of the CSI is not that straightforward. Depending on who computes the CSI, there are two CSI feedback schemes: (1) the implicit feedback (Figure \ref{Fig:imp_csi}), where the AP computes the CSI by estimating training sequences sent from STAs, and (2) the explicit one (Figure \ref{Fig:exp_csi}), where STAs calculate the CSI by estimating the training sequence sent from the AP, and then STAs feedback the calculated CSI to the AP. 

By assuming the reciprocity of up/down-link channels, the implicit feedback scheme produces less overheads compared to the explicit one. However, in a practical WLAN system, the channel and interference seen by the STAs are generally not the same as those seen by the AP due to their different transmitting/receiving filters and PHY paths. Therefore, the antenna calibration \cite{IEEEac} is usually needed to reduce the distortion if implicit feedback is adopted. 

The explicit feedback scheme, i.e., STAs feedback the CSI, provides higher CSI resolution, but also higher overhead. MAC control frames are usually extended to support the CSI feedback in the literature, while an explicit compressed Feedback (ECFB) scheme is introduced by IEEE 802.11ac to schedule and compress the volume of CSI feedback.

No matter which CSI feedback scheme is applied, the implicit one or the explicit one, the frequency of CSI feedback would significantly affect the network performance. It is because the frequent CSI feedback increases overheads, while the infrequent one results in the outdated CSI that leads to interference among parallel streams. Please refer to \cite{lou2013comparison} and \cite{gong2010training} for more details about implicit and explicit CSI feedback schemes. 

\subsection{The Scheduling Scheme}
Another key point for designing MU-MIMO MAC protocols is the scheduling scheme. It is used to select a group of STAs or frames for transmissions, which can optimize certain aspects of the system performance according to the specific grouping criteria. The design of the scheduling scheme can be divided into two parts: the scheduling in the uplink and downlink. The latter can be easily categorized by different scheduling algorithms (e.g., the destination based round-robin scheme or the frame based first-in first-out scheme), while the characterization of the former is not straightforward. The reason and categorization of the uplink scheduling schemes are as follows.

\subsubsection{Scheduling in the Uplink}
\begin{figure}[h!!!!!!!!]
\centering
\includegraphics[scale=1.1]{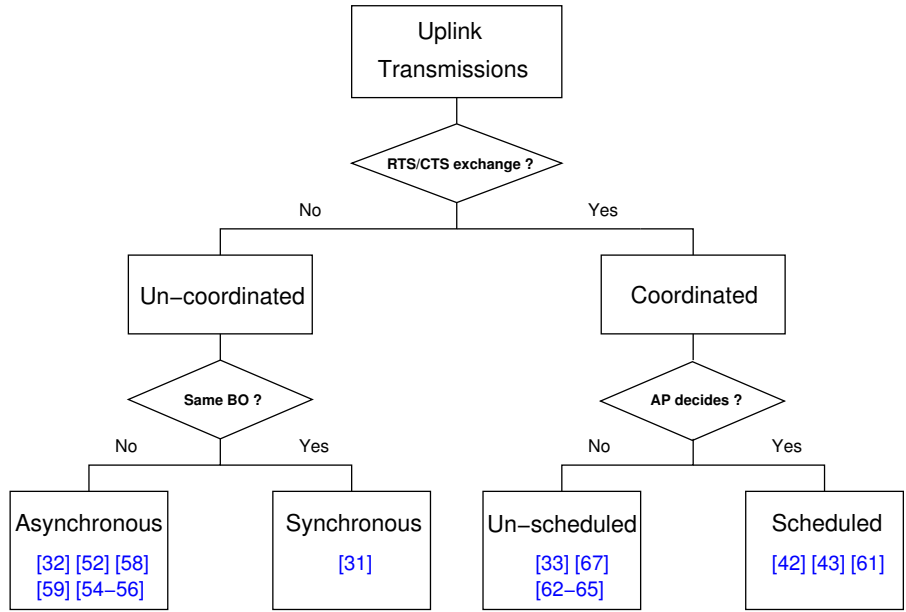}
\caption{Categories of uplink transmissions}
\label{Fig:up_coor_sch}
\end{figure}

In the uplink, it is very challenging to make a joint scheduling decision among spatially distributed STAs. As shown in Figure \ref{Fig:up_coor_sch}, only three of surveyed papers considered scheduling schemes, which will be described in details in the protocol review Section \ref{sec:survey}. 


\begin{figure}[h!!!!!!]
\begin{center}
\subfigure[Synchronous data transmissions]{\includegraphics[scale=0.5]{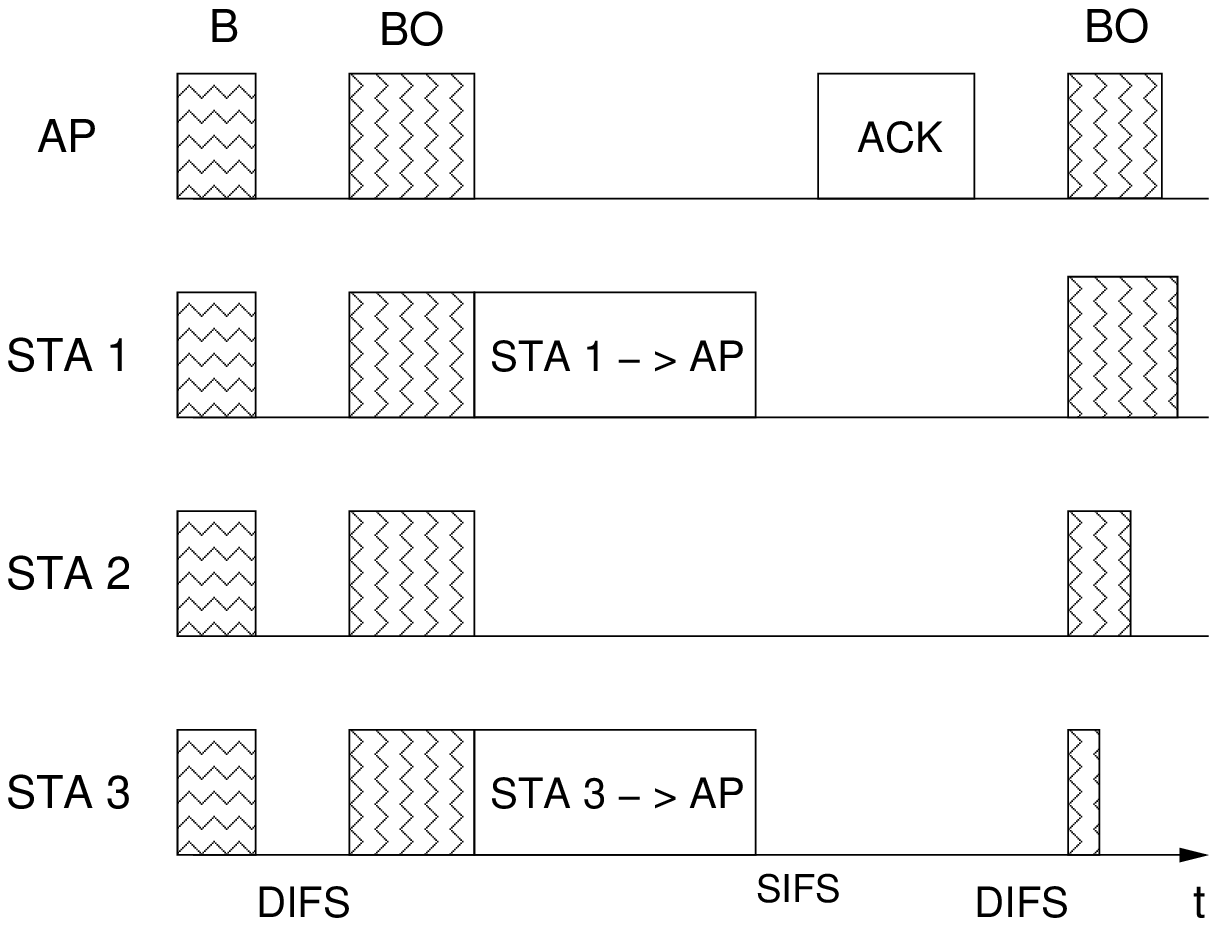}{\label{Fig:unmr1}}}
\subfigure[Asynchronous data transmissions]{\includegraphics[scale=0.5]{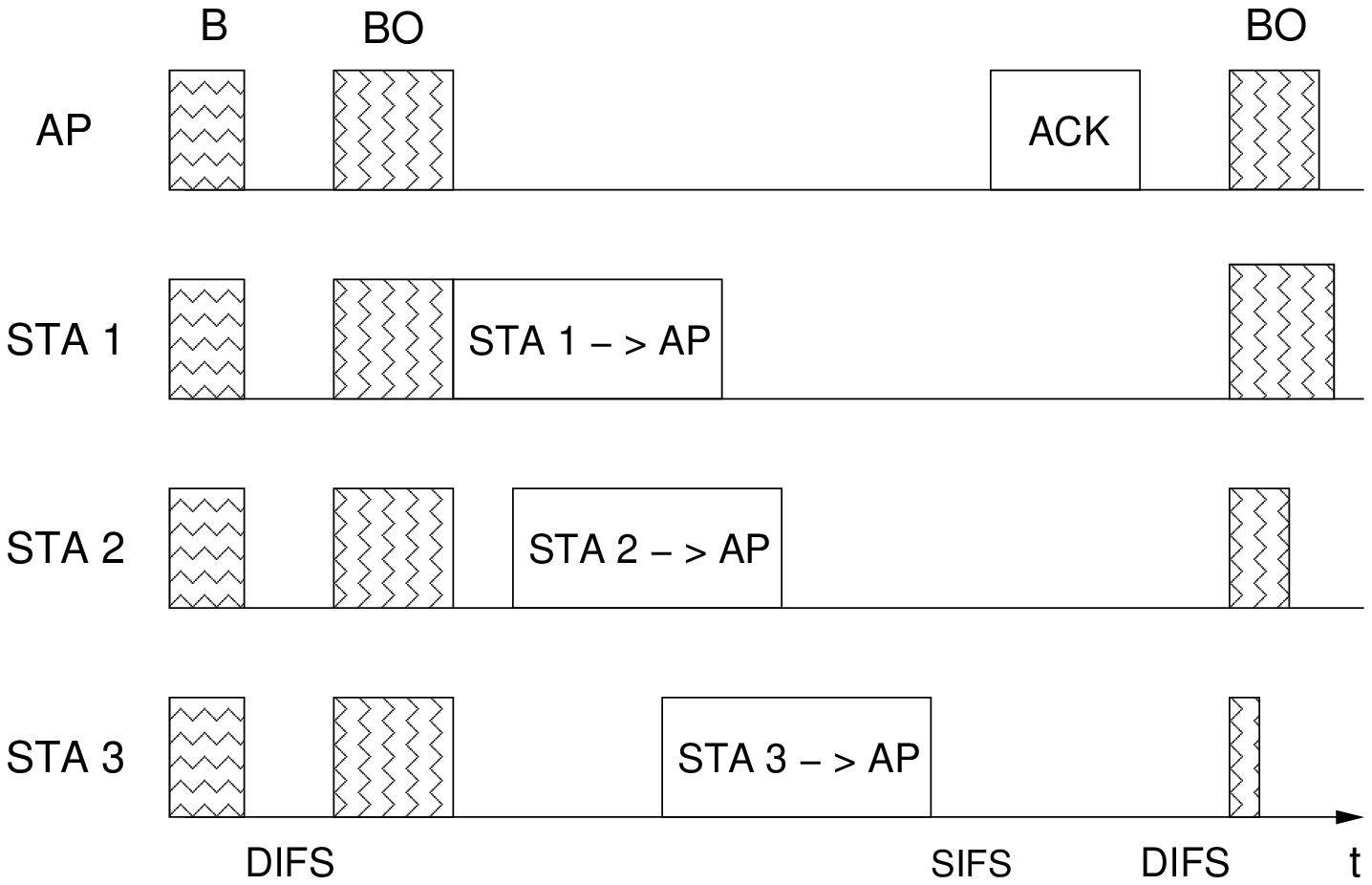}{\label{Fig:unmr2}}}
\caption{Un-coordinated uplink channel access}\label{Fig:uncoor}
\end{center}
\end{figure}

Depending on whether the RTS/CTS exchanging process is employed (i.e., whether the AP has played a coordinating role in exchanging control frames before transmitting data), uplink transmissions are categorized into the coordinated  and the un-coordinated ones. In the un-coordinated scenario, STAs utilize the random MAC mechanism to decide who will be allowed for transmissions, which have two cases: synchronous \cite{jhup} and asynchronous \cite{BabichC10} data transmissions, as shown in Figure \ref{Fig:uncoor}. The synchronous scheme lets multiple STAs that coincidentally choose the same BO to transmit data frames simultaneously, while the asynchronous one allows STAs to transmit frames along with other ongoing transmissions.

\begin{figure}[h!!!!!!!!]
\centering
\includegraphics[scale=0.5]{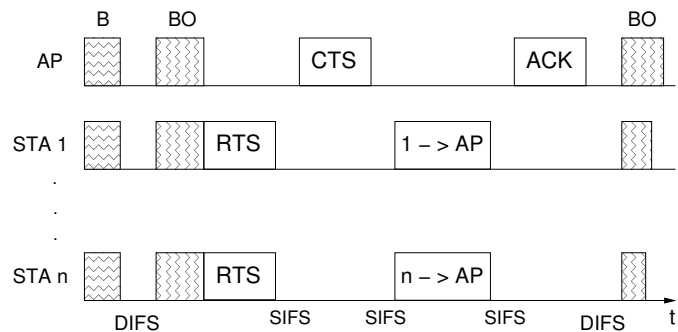}
\caption{Coordinated uplink channel access}
\label{Fig:cmpr}
\end{figure}

In the coordinated scenario, STAs utilize the MAC random mechanism to contend for the channel, while let the AP to decide who will be involved in the followed parallel transmissions. The coordinated uplink access scheme implies the involvement of the AP (as a coordinator) and the employment of RTS/CTS exchanges \cite{PZheng}. The AP extracts the interested information from RTSs sent by the contending STAs, and then makes scheduling decisions for simultaneous frame transmissions (i.e., the scheduled transmissions), or the AP just responds to the received RTSs to notify who have won the channel contention (i.e., the un-scheduled transmissions). A general example of the coordinated uplink access is shown in Figure \ref{Fig:cmpr}, which can account for both scheduled and un-scheduled cases depending on whether the CTS is extended to support scheduling.

Although the un-coordinated uplink channel access requires fewer modifications compared to the coordinated one, it is likely that the spatial domain will be underutilized. The reason is that the concurrent uplink access of the un-coordinated scheme is based on the randomness of the IEEE 802.11 backoff mechanism. In comparison, the coordinated uplink channel access lets the AP to mediate the uplink transmissions by either just notifying a group of STAs that successfully won the channel contention or making a scheduling decision that aims to optimize the system performance. Obviously, the coordinated scheme will introduce overheads (e.g., extra fields in RTS/CTS) that are needed for the AP to best exploit the spatial domain. 

\subsubsection{Scheduling in the Downlink}
Compared to the uplink, the AP plays a more direct role in the downlink scheduling, which can be classified into the packet based scheduling and the STA based one. In principle, the following downlink scheduling schemes can be applied to uplink transmissions. However, they are adopted in very few proposals due to the difficulty and overheads to make uplink scheduling decisions for spatially-distributed STAs.

The packet based scheduling algorithms utilize the packet queueing status at the AP as the scheduling metric to assemble multiple packets for MU-MIMO downlink transmissions. The relevant packet based scheduling algorithms include First-in First-out (FIFO) \cite{rldde,bellalta2012performance,liao2013uni,6566748}, weighted fair queue (WFQ, weight based on the priority or the type of packets) \cite{LCai,mxgac}, earliest deadline first (EDF, based on the delay or waiting time of packets), greedy \cite{zhang2010employing,hshca}, etc.

The STA based scheduling employs some criteria to identify a set of STAs for simultaneous downlink transmissions. These criteria include the channel state \cite{whuaj,tatae,EKar,zhao2009applying,hlimm}, spatial compatibility, fairness, etc. The pure channel condition based scheduling is also called opportunistic scheduling that singles out a set of STAs with best channel conditions by benefiting from the multi-user diversity. A similar concept is applied by the spatial compatibility based scheduling, which examines STAs' spatial correlations to minimize the interference. Neither of them has taken the fairness into account, which is considered by the round-robin scheduling.

A combination of the above mentioned scheduling schemes is usually considered. For example, the proportional fair scheduler usually considers a product of the channel capacity and fairness. However, as we can see from the above, the most commonly used scheduling schemes are FIFO and opportunistic, which are followed by greedy and WFQ. The reasons could be as follows. Comparing to cellular networks, (1) WLANs consist of fewer STAs, (2) WLAN STAs are more stable in terms of moving speeds and residing environments, and (3) the AP is usually less powerful than the base station.

\begin{figure}[h!!!!!!!!]
\centering
\includegraphics[scale=1.4]{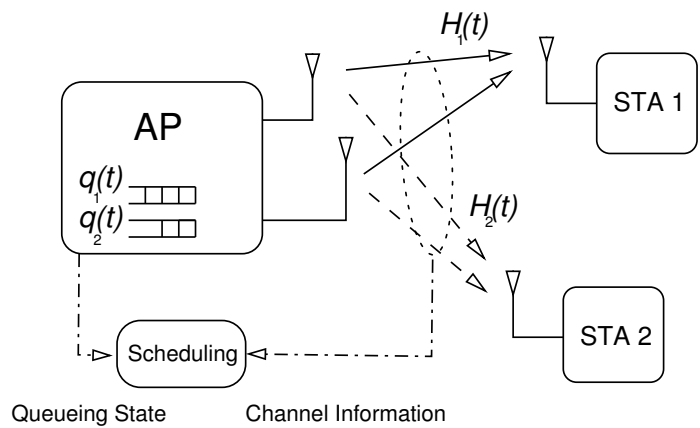}
\caption{Queueing and channel based cross-layer scheduling}
\label{Fig:cross}
\end{figure}

The combination of several criteria hints that parameters from different layers should be jointly considered \cite{CAnt}, which is the concept of cross-layer scheduling as shown in Figure \ref{Fig:cross}, where $q(t)$ and $H(t)$ represent the queueing and the channel states at the time $t$.

\subsubsection{Cross-layer Scheduling}
The cross-layer scheduling has the promise to achieve the optimal system performance by sharing and configuring parameters from different layers, such as the channel information at the PHY layer, the queueing state at the MAC layer and the routing information at the network layer. Unfortunately, the cross-layer scheduling remains far more complex than simply combining these parameters. The reason is that the interaction among the layers breaks the conventional Open System Interconnection (OSI) layered structures and creates tensions between the performance and the stability of systems, which could lead to unexpected consequences as the wireless network scales up \cite{vkawa}.

In general, together with the CSI acquisition and other layers' key parameters, the following points should be taken into account for the cross-layer design.

\begin{itemize}
\item{\textbf{System Complexity}}: As the cross-layer design breaks the conventional layered structure, the new wireless system could be incompatible with conventional ones. Besides, since the maintenance or upgrade of the cross-layer protocol is no longer isolated within each layer, any parameter changes must be carefully traced and coordinated \cite{vkawa}.
\item{\textbf{Design Constraints}}: Various data rates are usually applied to spatially distributed STAs, which may cause interferences from stronger signals to weaker ones in the downlink or the near-far effect in the uplink. Therefore, a power control or a data rate selection scheme needs to be considered with the MAC design. In addition, some QoS metrics, such as the average delay and the jitter, are traditionally not in line with the MAC focus (e.g., decreasing collisions and increasing throughput). Sometimes, maximizing throughput means sacrificing transmission opportunities of some low-rate STAs. Thus, the tradeoff of different performance metrics needs to be jointly considered and given different weights \cite{whuac} \cite{yu2013resource}. 
\end{itemize}

\section{Survey of MU-MIMO MAC Protocols for WLANs}\label{sec:survey}
In this section, we look into prominent MU-MIMO MAC proposals in the literature by focusing on the required features, performance gains, evaluation tools, as well as the key assumptions they made.

\subsection{MAC Proposals for The Uplink}

\subsubsection{Un-coordinated Channel Access}
\paragraph{Synchronous Data Transmissions}
Jin et al. in \cite{jhup} present a simple MU-MIMO MAC scheme that relies on the simultaneous transmissions from STAs. The MAC procedure is the same as illustrated in Figure \ref{Fig:unmr1}. The authors assume each STA has an orthogonal preamble so that the AP can estimate the channel coefficients and differentiate STAs. Once the AP knows the channel, it employs the ZF scheme to separate the received signals. The authors extend the Markov chain model proposed by Bianchi in \cite{840210} to analyze the performance of the proposed MU-MIMO scheme in saturated conditions. Compared with SU-MIMO, the numerical results show that the proposed MU-MIMO scheme obtains lower collision probability, shorter delay and higher throughput in the low SNR and small network conditions.

\paragraph{Asynchronous Data Transmissions}

Tan et al. in \cite{kutse} present a practical Spatial Multiple Access (SAM) scheme for WLANs. SAM relies on a distributed MAC scheme called Carrier Counting Multiple Access (CCMA) to allow asynchronous concurrent transmissions. A chain decoding technique to separate simultaneously received frames is adopted. Each STA maintains a transmission counter by detecting other STAs' frame preambles, and decides whether to contend for the channel. SAM is evaluated in SORA, a Software Defined Radio (SDR) platform developed by Microsoft \cite{sora}. Evaluation results show that the proposal can increase the throughput by $70\%$ over the default IEEE 802.11 DCF.

Babich et al. in \cite{BabichC10} develop a Markov chain analytical model for asynchronous Multi-Packet Reception (MPR), where a STA is allowed to transmit even if other STAs are already transmitting. More specifically, a STA is allowed to decrease its BO counter as long as the channel is empty or the sensed number of ongoing transmissions is below a threshold. A generic error correction code is assumed to protect data frames. The MAC procedure is shown in Figure \ref{Fig:unmr2}. Based on the results obtained from the presented theoretical model and simulations, the authors claim that the asynchronous MAC scheme can provide considerable performance gains compared to the synchronous one due to higher utilization of the channel.

Ettefagh et al. in \cite{ettefagh2011performance} present a cluster-based MAC protocol for MU-MIMO transmissions called CB-CSMA/CA. In CB-CSMA/CA, STAs are grouped in clusters, and the AP is assumed to have MPR capabilities. STAs belonging to the same cluster share a common backoff counter, which allows simultaneous transmissions from multiple STAs when the counter reaches zero. As a result of this, the channel access is competed among clusters. Simulation and analytic results show that CB-CSMA/CA outperforms a pure STA-based contention approach since it increases the chance that more than one STA can transmit at the same slot.

Mukhopadhyay et al. in \cite{mukhopadhyay2012acknowledgement} explore the ACK-delay problem that arises in asynchronous MPR. The ACK-delay problem refers to that, due to different transmission durations of asynchronous uplink transmissions, the delayed ACKs sent by the AP to those earlier-finished STAs can trigger STAs' ACK time-out counters, which could interrupt the ACK's transmission and degrade the network performance. Since Babich et al. in \cite{BabichC10} did not consider the ACK-delay problem, the authors in \cite{mukhopadhyay2012acknowledgement} propose to change the ACK-waiting STAs' backoff timer to be decreased only when the channel has been idle for DIFS. The simultaneous transmissions are assumed to be decodable by the AP, and a single data rate is also assumed. By comparing with Babich's and IEEE 802.11 standard schemes, the results show that the proposed one not only decreases the frame collision probability and the average delay, but also increases the throughput.

Lin et al. in \cite{6566751} propose a MIMO concurrent uplink transmission scheme called MIMO/CON, which can support both asynchronous and synchronous data transmissions. A compressive sensing technique \cite{candes2008introduction} is utilized to estimate CSI from multiple concurrently received preambles, and ZF is adopted to separate the data frames. A delayed packet decoding mechanism, namely, using partially retransmitted information to decode the collided frames, is devised to avoid the complete retransmission of all corrupted frames. A fixed frame length is assumed, and the optimal transmission probability is assumed to be known by STAs. Tan's CCMA \cite{kutse} is implemented to compare with the proposed MIMO/CON. The results show that CCMA outperforms MIMO/CON when the AP has fewer antennas, while MIMO/CON scales better as the number of antennas at the AP increases. 

Wu et al. in \cite{6807572} propose a throughput analytical model for asynchronous uplink transmissions, which is based on the Bianchi Markov chain model \cite{840210}. A beacon sent by the AP will announce the maximum number of STAs that are allowed to transmit in parallel. The MAC procedure is similar to the one described above, in \cite{kutse}. A fixed data rate is assumed and the network is saturated. Zero-forcing with successive interference cancellation (ZF-SIC) is employed to decode parallel data streams. By numerically deriving the contention window size and other network parameters, the uplink throughput is maximized. With those parameters, the authors derive a threshold of the number of antennas at the AP, which shows no throughput benefit can be achieved by adding more antennas. The reasons for that are: (1) the available transmission time decreases as the number of STAs involved in the parallel transmission increases; and (2) the collision probability increases as more antennas are employed at the AP. 

Kuo et al. in \cite{kuo2014leader} present a leader-based MAC protocol for uplink MU-MIMO transmissions. Similar to other asynchronous proposals, each STA counts the number of concurrent transmissions to decide if it can start a new transmission. In case there are multiple on-going transmissions, the proposed protocol requires them to finish at the same time. The main difference compared with previous proposals is that only the first transmission is randomly selected (i.e., a leader will follow the CSMA/CA rules). The remaining STAs will only start a concurrent transmission if they overheard a transmission from the leader. The STAs' scheduling is determined by a user matching solution developed in the paper aiming to both maximize the system throughput and fairness.

Table \ref{upmac} summarizes the main characteristics of the surveyed un-coordinated uplink MU-MIMO MAC protocols.

\begin{table*}[t!!!!!!!]
\caption{{ \bfseries Un-coordinated uplink MU-MIMO MAC protocols}}
\newsavebox{\Tablebox}
\begin{lrbox}{\tablebox}
 \rowcolors{0}{blue!10}{}
\renewcommand\arraystretch{3}
{\Huge
\begin{tabular}{|l|l|l|l|l|l|l|}
\hline
{\bfseries \Huge Remarks}  & {\bfseries \Huge Evaluation Tool}  & {\bfseries \Huge CSI Scheme} & {\bfseries \Huge MUD} & {\bfseries \Huge Key Assumption} & {\bfseries \Huge Scheduling}\\ \hline 
Jin \cite{jhup}, compare SU and MU-MIMO, 2008 & Analysis  & Implicit feedback & Zero forcing & Orthogonal preambles & - \\ \hline

Tan \cite{kutse}, carrier counting, 2009 & Testbed & Implicit feedback & Chain decoding & - & - \\ \hline

Babich \cite{BabichC10}, asynchronous MPR, 2010 & Analysis  & - & - & Code correction scheme & - \\ \hline %

Ettefagh \cite{ettefagh2011performance}, cluster-based MU-MIMO, 2011 & Simulation + Analysis  & Implicit feedback & - &  Ideal channel & -\\ \hline %

Mukhopadhyay \cite{mukhopadhyay2012acknowledgement}, ACK-aware MPR, 2012  & Simulation + Analysis  & - & - &  MPR frames decodable & - \\ \hline %

Lin \cite{6566751}, delay packet decoding, 2013 & Simulation + Testbed  & Compressive sensing & Zero forcing & Fixed frame length & - \\ \hline

Wu \cite{6807572}, throughput model, 2014 & Simulation + Analysis  & Implicit feedback & ZF-SIC & Fixed data rate & - \\ \hline %

Kuo \cite{kuo2014leader}, leader-based contention, 2014 & Simulation + Analysis  & Implicit feedback & ZF-SIC & Rayleigh fading & User matching \\ \hline %

\end{tabular}}
\label{upmac}
\end{lrbox}
\resizebox{1\textwidth}{!}{\usebox{\tablebox}}
\end{table*}

\subsubsection{Coordinated Channel Access}
\paragraph{Scheduled Data Transmissions}
Huang et al. in \cite{whuaj} present an MPR MAC protocol that utilizes CDMA to separate the compound frames. When the STA's backoff counter reaches zero, a function that considers the number of STAs in the network, the current channel state and the MPR capability of the AP is used to schedule STAs. The MAC procedure is illustrated in Figure \ref{Fig:cmpr}. The CSI is claimed to be obtained from the downlink transmission. Data frames are assumed to have the same length. The analysis is based on the Bianchi's model \cite{840210}, and the optimal transmission probability is obtained by one-dimensional search procedure, such as Bisection and Newton-Raphson method \cite{chong2013introduction}. Results obtained from the analytic model and simulations show that the proposed scheme reduces collisions and avoids considerable transmission errors. 

Tandai et al. in \cite{tatae} propose a synchronized access scheme coordinated by the AP. On receiving applying-RTSs (A-RTSs) from STAs, the AP responds with a pilot-requesting CTS (pR-CTS) to expect pilots. Based on the CSI estimated from the sequential pilots, the AP sends a Notifying-CTS (N-CTS) to inform the selected STAs for parallel transmissions. A unique subcarrier is assumed to be allocated to each STA to differentiate A-RTSs, and the MMSE decoder is adopted to separate the simultaneously received signals. According to the simulation results, the proposed scheme can reduce the overhead and increase the throughput.

Li et al. in \cite{li2014cuts} present a coordinated MAC protocol to improve Channel Utilization in both Time and Spatial domain (CUTS). In the considered approach, the AP first sends a grant packet to all STAs. Those backlogged STAs reply to the AP with a transmission request. Then, the AP analyzes all the information received, selects and signals those allowed STAs to transmit simultaneously. The main innovation is that the channel contention is done in the frequency domain. In CUTS, each STA selects one of the available subcarriers to transmit a predefined signal, which is used by the AP to identify the winning STAs. Both experimental and simulation results show the significant gains achieved by CUTS.

\paragraph{Un-scheduled Data Transmissions}
Zheng et al. in \cite{PZheng} propose a MU-MIMO MAC protocol called MPR-MAC, which extends CTS and ACK to accommodate multiple transmitters. The MPR-MAC procedure is the same as illustrated in Figure \ref{Fig:cmpr}, where the STAs that won the channel contention will transmit RTSs at the same time. The AP then replies with an extended CTS that grants concurrent transmissions to the requesting STAs. A set of orthogonal training sequences are assigned to STAs by the AP to facilitate the channel estimation. A Finite Alphabet (FA) based blind detection scheme is adopted for the frame separation. The network is assumed to be saturated, and all data frames have the same length. The throughput analytic model is based on Bianchi's work \cite{840210}, and the best transmission probability is obtained via numerical techniques (i.e., a derivation function). The authors claim that the throughput increases nearly linearly with the number of antennas at the AP. 

\begin{figure}[h!!!!!!!!]
\centering
\includegraphics[scale=0.5]{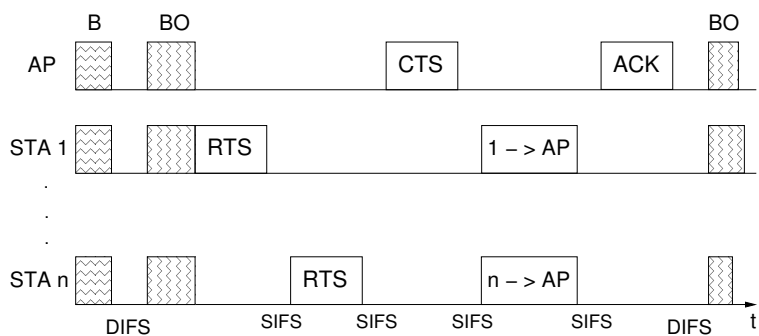}
\caption{TRRC: Two-Round RTS Contention}
\label{Fig:mpr2}
\end{figure}

Since the MPR-MAC follows the conventional IEEE DCF access scheme, the probability of more than one STA choosing the same random BO is low. In order to increase the number of parallel transmissions, an enhancement called Two-Round RTS Contention (TRRC, Figure \ref{Fig:mpr2}) is also proposed in \cite{PZheng}. Compared to MPR-MAC, TRRC has two RTS contention rounds. Namely, instead of sending a CTS after the first RTS round, the AP waits for an extra round to recruit more RTSs. The results show that TRRC obtains a further $7\%$ throughput increase. 

Barghi et al. in \cite{barghi2011mimo} present an MPR-aware MAC protocol for receiving two concurrent frames by introducing a waiting time window ($t_\text{w}$) at the AP. Specifically, when the AP receives a first RTS, it will wait for a time period of $t_\text{w}$ to recruit a second RTS for double-frame receptions. CTS and ACK are extended with an extra address field to accommodate two STAs. A space-time code scheme is adopted to detect multiple frames. The channel is assumed to be error-free and channel coefficients are assumed to be known. Based on the obtained results, the authors claim that, by widening $t_\text{w}$, (1) the probability of double-frame transmissions increases, while the probability of collision (i.e., the probability that the number of transmitted RTSs is more than two) increases as well; and (2) the performance of the proposed MPR-aware MAC scheme improves significantly compared to that of the IEEE 802.11 standard one.

\begin{figure}[h!!!!!!]
\begin{center}
\subfigure[Without timeout]{\includegraphics[scale=0.5]{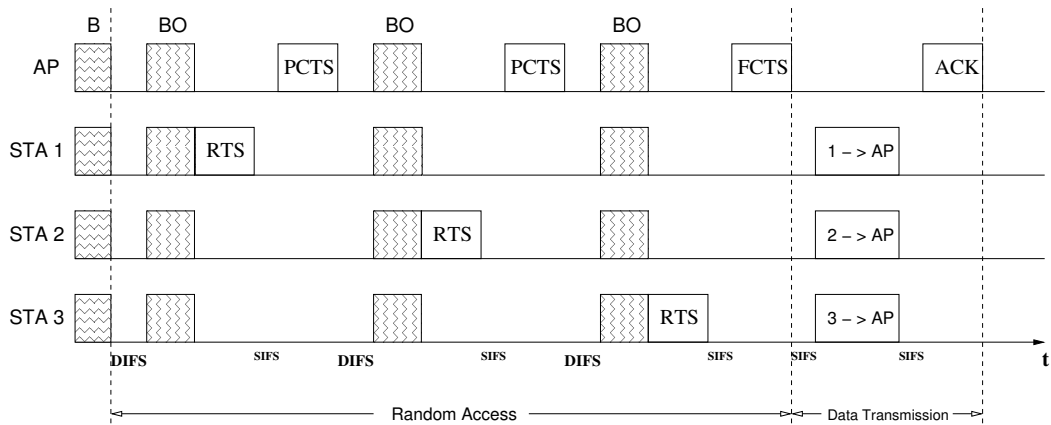}{\label{Fig:mr1}}} \\
\subfigure[With timeout]{\includegraphics[scale=0.5]{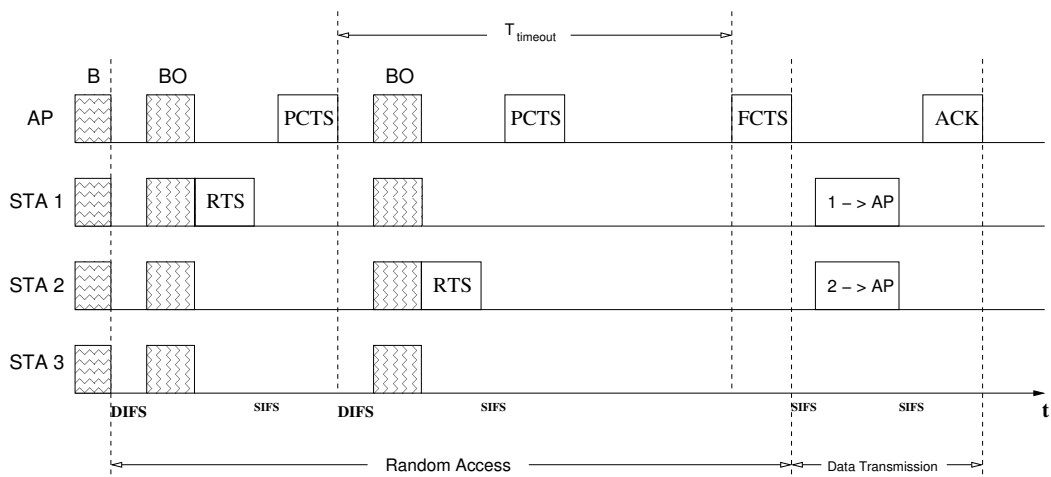}{\label{Fig:mr2}}}
\caption{Two-round MAC procedures, $M_{\text{random}} = 3$}\label{Fig:mrandom}
\end{center}
\end{figure}

Zhou et al. in \cite{SZhou} propose a two-round channel contention mechanism, which divides the MAC procedure into two parts. The two parts, namely, the random access and the data transmission, are illustrated in Figure \ref{Fig:mrandom}. The random access finishes as soon as the AP receives $M_{\text{random}}$ (the maximum number of STAs that can transmit simultaneously) successful RTSs, and then the data transmission starts. In the random access part, the AP delivers two types of CTSs: Pending CTS (PCTS) and Final CTS (FCTS). The former responds to RTS and the latter notifies all STAs about the start of data transmissions. Those STAs, who have sent RTSs within a predefined time threshold ($T_{\text{timeout}}$), will transmit simultaneously. If the number of contending users is less than $M_{\text{random}}$, the $T_{\text{timeout}}$ will trigger data transmissions as well. The AP obtains the CSI from RTSs, and utilizes the MMSE detector to separate STAs' signals. Data frames are assumed to be of fixed length. Both simulation and analytic results show that the two-round contention scheme outperforms the IEEE 802.11 single-round one in terms of throughput and delay.


A similar work with two contention rounds is presented in \cite{rlidu}. Compared to \cite{SZhou}, \cite{rlidu} devises a shorter second contention round, where a single message is used to reply all successfully received RTSs. In addition, a special focus is placed on $2$-nd round Contention Window ($CW_{\text{2nd}}$), a parameter making the length of the second contention round elastic. By evaluating the proposal in simulations, a set of optimal $CW_{\text{2nd}}$ values that can obtain the highest system performance are identified.

Zhang in \cite{yjzmr} further extends two contention rounds to multiple rounds, which give STAs more opportunities to compete for the channel using a threshold derived from an optimal stopping algorithm \cite{chow1971great}. Meanwhile, an auto fall-back to single-round scheme is also proposed in case the traffic is low and the single-round scheme can provide higher throughput. Frame arrivals are assumed to follow the Poisson distribution. Results obtained from simulations and the analytical model show the multi-round contention can increase the channel utilization rate in a small to moderate network.

\begin{figure}[h!!!!!!!!]
\centering
\includegraphics[scale=0.6]{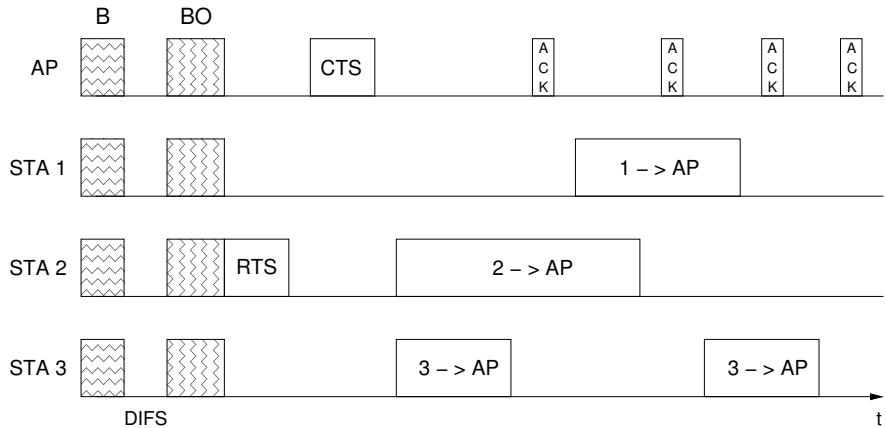}
\caption{Asynchronous data transmissions with one MPR vacancy}
\label{Fig:coorasy}
\end{figure}

Jung et al. in \cite{6302115} present a coordinated uplink MPR scheme, which extends the work in \cite{BabichC10} to allow both synchronous and asynchronous transmissions by employing an additional feedback channel. The proposed MAC procedure is shown in Figure \ref{Fig:coorasy}. On receiving an RTS from STA$2$, the AP replies with a CTS that includes the MPR vacancy (the remaining space for parallel uplink transmissions). STAs who overhear the MPR vacancy will compete for the channel to transmit along with STA$2$. Once a STA finishes transmitting ahead of the other one, the AP immediately sends an ACK with the updated MPR vacancy information through the additional channel to allow other STAs to compete for the newly available MPR space. The authors assume an orthogonal training sequence is included in the preamble of each frame for estimating the channel. Based on results obtained from the Markov chain analytic model and simulations, the authors claim that the proposed scheme achieves higher channel efficiency in scenarios where the frame size and transmission rates are dynamically varying.

Table \ref{unpmac} summarizes the main characteristics of the surveyed coordinated uplink MU-MIMO MAC protocols.

\begin{table*}[t!!!!!!!]
\caption{{ \bfseries Coordinated Uplink MU-MIMO MAC Protocols}}
\begin{lrbox}{\tablebox}
 \rowcolors{0}{blue!10}{}
\renewcommand\arraystretch{3}
{\Huge
\begin{tabular}{|l|l|l|l|l|l|l|}
\hline
{\bfseries \Huge Remarks } & {\bfseries \Huge Evaluation Tool} & {\bfseries \Huge CSI Scheme} & {\bfseries \Huge MUD} & {\bfseries \Huge Key Assumption} & {\bfseries \Huge Scheduling}\\ \hline
Huang \cite{whuaj}, SNR based MPR, 2008 & Simulation + Analysis & Downlink estimation & CDMA & Fixed data length & Optimal SNR  \\ \hline

Tandai \cite{tatae}, TDMA signalling, 2009 & Simulation & Implicit feedback & MMSE  & Unique subcarrier & Best CSI  \\ \hline

Li \cite{li2014cuts}, subcarrier contention, 2014 & Simulation + Testbed & Implicit feedback & - & Error-free channel & Subcarrier  \\ \hline

Zheng \cite{PZheng}, DCF based MPR, 2006& Analysis & Implicit feedback & Blind detection  & Fixed data length  & -  \\ \hline


Barghi \cite{barghi2011mimo}, MPR-aware MAC, 2011& Simulation + Analysis  & - & STC & Perfect channel & -  \\ \hline %

Zhou \cite{SZhou}, two-round contentions, 2010& Simulation + Analysis & Implicit feedback & MMSE  & Fixed data length & -  \\ \hline

Liao \cite{rlidu}, elastic $2$-nd round, 2012& Simulation & Implicit feedback & -  & Error-free channel & -  \\ \hline

Zhang \cite{yjzmr}, multi-round contentions, 2010& Simulation + Analysis & - & -  & Poisson arrivals & -  \\ \hline


Jung \cite{6302115}, asynchronous MPR, 2012& Simulation + Analysis & Implicit feedback & - & Extra ACK channel & -  \\ \hline

\end{tabular}}
\label{unpmac}
\end{lrbox}
\resizebox{1\textwidth}{!}{\usebox{\tablebox}}
\end{table*}

\subsubsection{Discussions on Uplink MU-MIMO MAC Proposals}
Here, we conclude this subsection by providing highlights, open aspects and next steps after reviewing the uplink MU-MIMO MAC proposals in the literature.

	\paragraph{Highlights}

	\begin{itemize}
		\item In the un-coordinated category, by simply considering the number of summarized papers, the asynchronous access has got more attention than the synchronous one. This is because, in the synchronous case, we can only benefit from the simultaneous multi-frame transmissions when several STAs end their backoff at the same time. One point that seems not fully solved by asynchronous solutions is how reliable the distributed counter of concurrent transmissions at each node is, as which may negatively affect the network performance by causing collisions. In addition, it is unclear how such carrier sense should be implemented, as the nodes need to identify the exact number of on-going transmissions.
	    \item In the coordinated category, the AP may schedule (scheduled) or not schedule (unscheduled) what STAs to transmit. It is surprising to see that there are more unscheduled proposals. The reason could be (1) the difficulty and overheads behind the uplink scheduling, and (2) simply keeping the "decentralized" access philosophy for WLANs. 
	\end{itemize}

\paragraph{Open aspects}

	\begin{itemize}
		\item Comparison between different proposals is still difficult because of the different considered scenarios (e.g., channel models, CSI assumptions and STAs placement). Therefore, an open challenge is to present a comparative analysis framework to benchmark the most relevant and promising proposals.
		\item Most of the uplink proposals assume implicit CSI. However, calibration problems, due to the presence of heterogeneous devices and channel conditions, make the implicit CSI difficult to be implemented.
		\item It is likely that, in the next generation 802.11 standard, an extended RTS/CTS exchange will be used to signal and protect MU-MIMO frames, which would mean the AP-coordinated approach will be adopted. Solutions to further reduce the handshaking overheads must be considered.
	\end{itemize}

		\paragraph{Next steps} 

	\begin{itemize}
		\item The IEEE 802.11ax study group is discussing to include uplink MU-MIMO \cite{tnup}\cite{zlup}. At the current stage, it is not clear yet how such mechanism will be implemented. In our perspective, uplink MU-MIMO would follow a coordinated and scheduled approach controlled by the AP. This judgement is made based on the following two reasons. First, the AP is playing a central role in collecting CSI from all STAs for downlink MU-MIMO transmissions in 802.11ac. We believe this explicit CSI feedback mechanism will be kept in 802.11ax for the backward compatibility. Additionally, by playing a central role, the AP can ask for buffer and other information from STAs, enabling it to make the best scheduling decision in the dense scenario. 
	  	\item Regarding the CSI of the uplink, in our opinion, 802.11ax could extend and reuse the explicit CSI scheme periodically conducted by 802.11ac for the downlink MU-MIMO transmissions. For example, the AP can estimate the uplink channel coefficients from the same packets that carry the downlink channel coefficients.
	  	\item 802.11ax should place STAs in virtual groups, and poll the STAs in each group based on the CSI information as well as the estimation of the buffer occupancy of the STAs to reduce overheads.
		%
	\end{itemize}
	

\subsection{MAC Proposals for The Downlink}
\begin{figure}[h!!!!!!!!]
\centering
\includegraphics[scale=0.6]{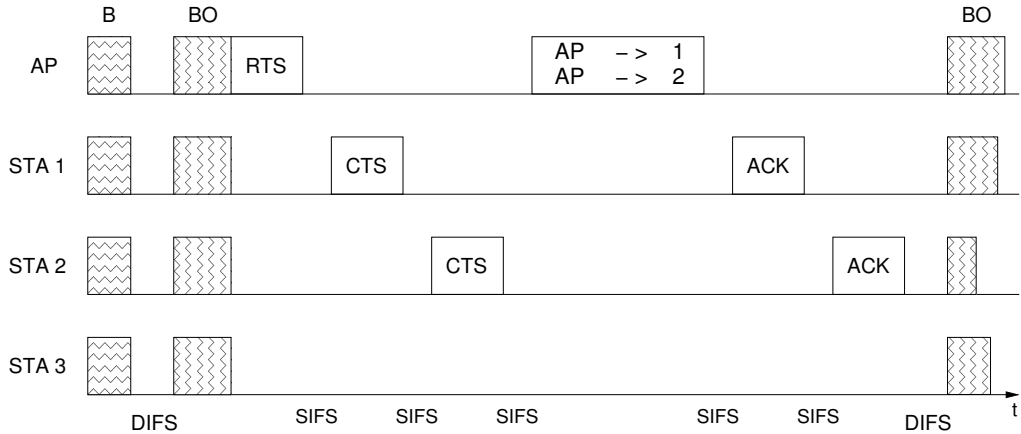}
\caption{A successful downlink MU-MIMO transmission}
\label{Fig:ss1}
\end{figure}

A commonly used MAC procedure for MU-MIMO downlink transmissions is illustrated in Figure \ref{Fig:ss1}. The AP firstly sends out a modified RTS containing a group of targeted STAs. On receiving the RTS, those listed STAs estimate the channel, integrate the CSI into the extended CTS and send it back. As soon as the AP receives all successful CTSs, it precodes the outgoing frames based on the feedback CSI. 

\subsubsection{Downlink MU-MIMO MAC Proposals}

Cai et al. in \cite{LCai} propose a distributed MU-MIMO MAC protocol with extended RTS/CTS frames. The CSI is obtained from the RTS/CTS exchange. An additive white Gaussian noise (AWGN) channel is assumed, and a leakage-based precoding scheme is utilized to cancel interference. The authors adopt a queue based scheduling scheme, which prioritizes frames with the longer waiting time in the AP buffer. The results derived from the queue-based unsaturated model \cite{cai2006voice} and simulations show the proposed multi-user MAC substantially outperforms the single-user one.

Kartsakli et al. in \cite{EKar} propose four multi-user scheduling schemes for concurrent frame transmissions, namely, MU-Basic, MU-Deterministic, MU-Threshold Selective, and MU-Probability. The opportunistic beamforming that selects STAs with the highest signal-to-interference-plus-noise ratio (SINR) for each randomly generated beam is utilized. The CSI is fedback by STAs during the channel contention phase. A block fading channel is assumed, which means the channel remains constant during a frame transmission time. Based on simulation results, the authors argue that the proposed schemes achieve notable gains against the single-user case, although there is still considerable space for improvements compared to the theoretical capacity.


Gong et al. in \cite{mxgac} propose a modified CSMA/CA protocol with three different ACK-replying mechanisms, namely, the polled ACK response, the scheduled ACK response and the failed ACK recovery. A weighted queueing mechanism that associates ACs with the value of contention window is also proposed to address the fairness concern. The MMSE precoding scheme is adopted and the CSI is assumed to be known. The simulation results show that the proposed protocol provides a considerable performance improvement against the IEEE 802.11n beamforming based approach when the SNR is high.

Zhang et al. in \cite{zhang2010employing} present a One-Sender-Multiple-Receiver (OSMR) transmission scheme for WLANs. The authors firstly implement an OSMR prototype using Universal Software Radio Peripheral (USRP) \cite{usrp} to explore the feasibility of OSMR at the PHY level. Then, based on the study of the OSMR PHY characteristics, they modify the RTS/CTS frames to support the channel estimation. Simulations of the proposed extension are conducted at the MAC level. A greedy scheduling algorithm is used to transmit as many frames as possible in a TXOP with the urgent frames being prioritized over the normal ones. The ZF precoding scheme is adopted, while a flat fading channel is assumed. The simulation results show that a significant performance gain is achieved by employing OSMR transmissions.

Liao et al. in \cite{rldde} present a MAC protocol for downlink MU-MIMO transmissions, where frames are scheduled to each STA by FIFO. The CSI is obtained through estimating the training sequence included in the CTS preamble. The channel is assumed to be error-free and independently fade from frame to frame, which creates independent channels from the AP to STAs. Simulation results show that a significant throughput gain is obtained by exploiting the spatial domain of the channel.

Bellalta et al. in \cite{bellalta2012performance} explore the performance of downlink MU-MIMO with the packet aggregation, and look into the interplay between the buffer size and the number of antennas at the AP. The impact of these two parameters on the system throughput and packet delay is also investigated. The proposed scheduler tries to maximize the number of transmitted packets in each stream, and also balance the duration of all streams. RTS/CTS control frames are modified for the CSI estimation, and ZF is used to null the interference. The analytical model is based on the work presented in \cite{bellalta2009space}, assuming packets' Poisson arrival in a non-saturated network. Both simulation and analytical results show the performance of the proposed system is close to optimal.

Redieteab et al. in \cite{redieteab2012mu} investigate three different transmission schemes in a PHY and MAC cross-layer platform, which are SU-MIMO, MU-MIMO with multi-user interference and MU-MIMO without multi-user interference. An IEEE 802.11ac channel model \cite{11acChannel} is utilized to emulate channel variations. The MAC layer is made to be compliant with the 802.11ac specification draft, and the amendment defined ECFB protocol is assumed to be employed to obtain the CSI. The ZF channel-inversion precoding scheme is used to decode frames. Based on the simulation results, the authors conclude that multi-user interference has important effects on MU-MIMO transmissions, e.g., it results in less throughput. Therefore, an automatic switching algorithm between SU-MIMO and MU-MIMO is suggested by the authors.

Cha et al. in \cite{jcpc} compare the performance of downlink MU-MIMO to Space Time Block Coding. The ZF precoder is utilized, and the CSI is obtained at the AP by receivers' feedback. A Rayeigh fading and error-free channel is assumed. The results show that the downlink MU-MIMO scheme produces a higher throughput than the STBC one if transmitted frames are of similar length, while the results reverse in a fast-varying channel due to the high overheads of the CSI feedback .

Balan et al. in \cite{hbaem} implement a distributed MU-MIMO system that consists of several multi-antenna APs, which are connected and assumed to be synchronous by a coordinating server. The authors employ Zero Forcing Beamforming (ZFBF) for the frame separation. The CSI is acquired from uplink pilot symbols. Blind Interference Alignment (BIA) is used when the CSI is unavailable. The system is evaluated in the Wireless Open-Access Research Platform (WARP) platform \cite{warp}. The experimental results show that the presented MU-MIMO system can achieve high data rates and approach the theoretical maximum throughput.

Zhu et al. in \cite{czme} investigate the required modifications for TXOP to support multi-user transmissions. The proposed scheme, called multi-user TXOP (MU-TXOP), enables a STA whose AC won the TXOP to share the transmission period with MPDUs of other ACs. The authors assume all STAs can be grouped for multi-user downlink transmissions. Simulation results show that the proposed scheme not only obtains a higher throughput, but is also more fair compared to the conventional one.

Ji et al. in \cite{ji2014cooperative} present a cooperative transmission scheme that addresses the redundant Network Allocation Vector (NAV) setting and the outdated SINR problems. The redundant NAV setting usually occurs at STAs located in the overlapped areas of neighboring WLANs, while the outdated SINR problem is caused due to the delay between the channel estimation and the data transmission. The authors utilize reserved bits in control frames to announce the last frame in a transmission to synchronize the NAV setting, and employ STAs' ACKs to re-estimate and correct the SINR. The analysis model assumes frames that arrive to STAs to follow the Poisson process. The results obtained from the model and simulations show the enhanced scheme can achieve noticeable performance gains compared to the sampled one.

Table \ref{downmac} summarizes the main characteristics of the surveyed downlink MU-MIMO MAC protocols.

\begin{table*}[t!!!!!!!]
\caption{{ \bfseries Downlink MU-MIMO MAC Protocols}}
\newsavebox{\TaBlebox}
\begin{lrbox}{\tablebox}
 \rowcolors{0}{blue!10}{}
\renewcommand\arraystretch{3}
{\Huge
\begin{tabular}{|l|l|l|l|l|l|l|}
\hline
{\bfseries \Huge Remarks} & {\bfseries \Huge Evaluation Tool} & {\bfseries \Huge CSI Scheme} & {\bfseries \Huge MUIC} & {\bfseries \Huge Key Assumption} & {\bfseries \Huge Scheduling}\\ \hline

Cai \cite{LCai}, reduce AP-bottleneck effect, 2008& Simulation + Analysis & Explicit feedback & Leakage coding & AWGN channel & Priority queue   \\ \hline

Kartsakli \cite{EKar}, $4$ scheduling schemes, 2009& Simulation  & Explicit feedback & Beamforming & Block fading & Highest SINR  \\ \hline

Gong \cite{mxgac}, ACK-replying schemes, 2010& Simulation  & - & MMSE & Assume CSI known & Weighted queue  \\ \hline

Zhang \cite{zhang2010employing}, OSMR, 2010& Simulation + Testbed  & Explicit feedback & Zero Forcing & Flat fading & Greedy  \\ \hline

Liao \cite{rldde}, throughput and delay gain, 2011& Simulation & Implicit feedback & - & Error-free channel & Per-STA FIFO  \\ \hline

Bellalta \cite{bellalta2012performance}, packet aggregation, 2012& Simulation + Analysis & Explicit feedback & Zero Forcing & Poisson arrival & Per-STA FIFO  \\ \hline

Redieteab \cite{redieteab2012mu}, PHY+MAC platform, 2012& Simulation  & Explicit feedback & Zero Forcing & - & -  \\ \hline

Cha \cite{jcpc}, STBC \& MU-MIMO, 2012& Analysis  & Explicit feedback & Zero Forcing & Error-free channel & -  \\ \hline

Balan \cite{hbaem}, multi-AP system, 2012 & Simulation + Testbed  & Implicit feedback & ZFBF, BIA & Phase synchronous & - \\ \hline

Zhu \cite{czme}, multi-user TXOP, 2012 & Simulation  & - & - & All STAs groupable & - \\ \hline

Ji \cite{ji2014cooperative}, outdated NAV and SINR, 2014 & Simulation + Analysis & Explicit feedback & - & Poisson arrival & - \\ \hline



\end{tabular}}
\label{downmac}
\end{lrbox}
\resizebox{1\textwidth}{!}{\usebox{\tablebox}}
\end{table*}

\subsubsection{Discussions on Downlink MU-MIMO MAC Proposals}


	\paragraph{Highlights}

	\begin{itemize}
	
		\item Regarding the CSI acquisition, some assume the channel reciprocity \cite{rldde}\cite{hbaem}, which allows the AP to obtain the CSI directly from frames transmitted by the STAs, while most papers adopt the explicit way, though none of them follow the approach defined in IEEE 802.11ac. However, it is worth to point out that no papers discuss or justify whether a SIFS interval is long enough for STAs to process received preambles and estimate the channel.
		
		
		\item In terms of the MUIC scheme, the Zero Forcing approach is the most common one due to its simplicity and efficiency.
			

	\end{itemize}

	\paragraph{Open aspects}

	\begin{itemize}
		\item Performance comparisons of different solutions, including the approach considered by IEEE 802.11ac, are still missing. It is of special interest to compare if it is better to obtain the CSI for each specific transmission, which guarantees the CSI is fresh, or only periodically, at the risk that the CSI may be outdated, hence creating higher interference between the different spatial streams. First works on this direction can be found in \cite{6328529}\cite{liao2013performance}, though more efforts are still required.
		\item In addition, analytical models on understanding the interactions between the CSI state, the number of active users, the buffer size, and other specific mechanisms such as packet aggregation, need more exploration. A first work is presented in \cite{bellalta2012performance}, where the authors evaluate the impact of the finite buffer size with random arrivals at the AP in a donwlink MU-MIMO system.
	\end{itemize}

	\paragraph{Next steps} Since the IEEE 802.11ac amendment has already been standardized, research on downlink MU-MIMO should focus on how the 802.11ac approach can be extended or optimized towards the next amendment-IEEE 802.11ax. Next steps could emphasize on tuning the downlink MU-MIMO mechanisms of 802.11ac, such as the CSI acquisition process (e.g., the rate of requests and the set of sampled nodes), scheduling algorithms considering both the instantaneous traffic and QoS requirements, and the integration with other IEEE 802.11 mechanisms (e.g., multi-cast video communications, as defined by IEEE 802.11aa \cite{maraslis2012ieee}. 

\subsection{MAC Proposals for Integrated Up/down-link}
Only a few works have considered both MU-MIMO uplink and downlink transmissions in a single MAC protocol.

\subsubsection{Integrated Up/down-link MU-MIMO MAC Proposals}

\begin{figure}[h!!!!!!!!]
\centering
\includegraphics[scale=0.5]{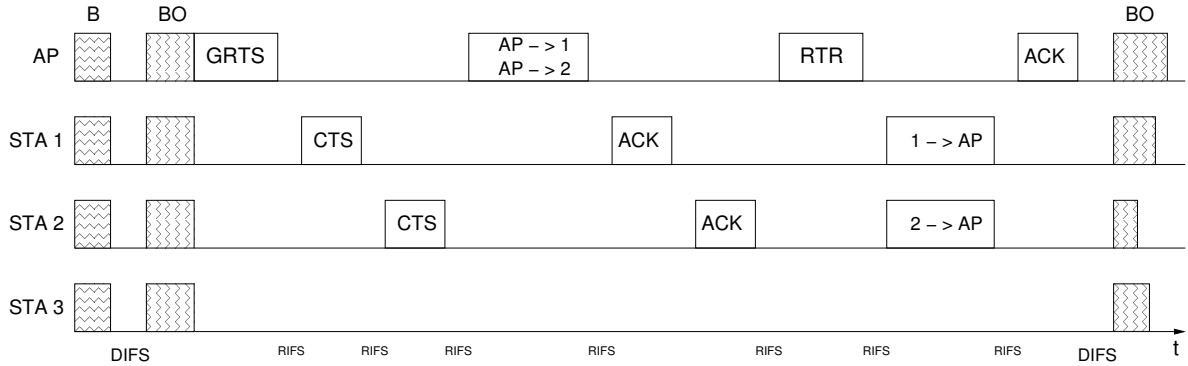}
\caption{PF-MAC: Per-flow MAC}
\label{Fig:back}
\end{figure}

Kim et al. in \cite{tkimm} devise a down/up-link back-to-back transmission scheme to synchronize STAs. The scheme is called Per-flow MAC (PF-MAC, Figure \ref{Fig:back}), where RIFS stands for Reduced Inter Frame Space. The AP first sends a Group-RTS (GRTS) that includes a list of STA addresses to initiate the downlink transmission. As soon as the AP received expected CTSs, it sends data frames to the listed STAs. The STAs who received frames then send back ACKs sequentially. Through the downlink transmission and a Ready to Receive (RTR) frame from the AP, STAs are synchronized for the parallel uplink transmission. The CSI is estimated from uplink frames. The limitation of the proposal is that the uplink transmission can only be started by the downlink one, which may not be desirable in some scenarios where the uplink access is urgent. Note that the proposal is just a conceptual model without any simulation or analytic results.

Zhao et al. in \cite{zhao2009applying} propose an opportunistic MU-MIMO MAC protocol for Multi-channel Multi-radio WLANs. In the downlink, a Group RTS is used to signal the selected STAs, which reply sequentially with a CTS before the AP transmits to them simultaneously. In the uplink, after the RTS/CTS exchange, the STAs are signalled to start transmissions using a Group CTS message. Note that both uplink and downlink transmissions can be done simultaneously as they use different channels. A common control channel is used to exchange information about the packet status and the set of available channels for next transmissions. The CSI is assumed available at the AP. The results show the proposal is able to outperform other multi-channel MAC schemes.

Li et al. in \cite{hlimm} propose a Multi-user MAC (MU-MAC) protocol, which supports Multi-Packet Transmission (MPT) in the downlink and multiple control packets (e.g., CTSs or ACKs) reception in the uplink. OFDM preambles are utilized to facilitate the CSI extraction from the simultaneously received control frames. The frame errors are assumed to come from collisions only, and all frames are transmitted with the same rate. The scheduling scheme jointly considers the CSI history, the AP's queueing state and the frames' application categories. The analytic model is based on and extended from \cite{LCai}. By observing the results, the authors claim that MU-MAC outperforms MPT-only MAC in terms of the maximum number of supported STAs, and this gain will further increase as the AP employs more transmitting antennas.

Shen et al. in \cite{hshca} present a High Throughput MIMO (HT-MIMO) MAC protocol that utilizes frequency signatures to differentiate simultaneously received control frames. HT-MIMO works in the PCF mode, hence both uplink and downlink transmissions can only be initiated by the AP. The CSI is obtained by a channel measurement method. The uplink and downlink channels are assumed to be symmetrical. A greedy scheduling algorithm is adopted with the consideration of fairness and the queue occupancy. By comparing with Cai's proposal \cite{LCai}, the analysis results show that the HT-MIMO MAC outperforms the 802.11n SU-MIMO and Cai's MU-MIMO schemes.

Liao et al. in \cite{liao2013uni} propose a unified MU-MIMO MAC protocol (Uni-MUMAC) for IEEE 802.11ac WLANs by integrating both uplink and downlink enhancements. An analytic model is developed based on \cite{840210}. The implicit CSI acquisition scheme is adopted by assuming the channel reciprocity. Through adaptively tuning $CW_{\text{2nd}}$ (a parameter controls the $2$nd uplink contention), and comparing with Li's proposal \cite{hlimm}, the authors conclude that (1) Uni-MUMAC performs well in both the traditional downlink-dominant and the emerging down/up-link balanced traffic scenarios; (2) one-side enhancement (e.g., the uplink enhancement) does not bring the same benefits to the other side.


Yun et al. in \cite{6566748} present a multi-point to multi-point MIMO system, where the uplink multiplexing is implemented in the SORA platform, while the downlink is implemented in the USRP platform. Multiple APs are coordinated by a controller, and connected via an Ethernet cable. A leader concept is adopted for both the uplink and  downlink medium access. A trigger frame that includes co-senders' addresses is sent out by the leader who first won the channel contention. Then, the co-senders transmit preambles sequentially as specified in the trigger frame for the CSI estimation. The downlink scheduling is based on the packet arrival time and the length of waiting time in the queue, while the co-senders' selection in the uplink is randomly made. ZF is employed in both the uplink and downlink for frames' de/pre-coding.

Table \ref{updownmac} summarizes the main characteristics of the Up/down-link MU-MIMO MAC protocols.

\begin{table*}[t!!!!!!!]
\caption{{ \bfseries Integrated Up/down-link MU-MIMO MAC Protocols}}
\begin{lrbox}{\tablebox}
 \rowcolors{0}{blue!10}{}
\renewcommand\arraystretch{3}
{\Huge
\begin{tabular}{|l|l|l|l|l|l|l|}
\hline
{\bfseries \Huge Remarks} & {\bfseries \Huge Evaluation Tool}  & {\bfseries \Huge CSI Scheme} & {\bfseries \Huge MUD \& MUIC} & {\bfseries \Huge Key Assumption} & {\bfseries \Huge Scheduling} \\ \hline

Kim \cite{tkimm}, back-to-back transmissions, 2008& Conceptual model  & Implicit feedback & - & - & -  \\ \hline

Zhao \cite{zhao2009applying}, multi-radio WLANs, 2009& Simulation & - & - & CSI known & Opportunistic  \\ \hline

Li \cite{hlimm}, MU-MAC, 2010& Simulation + Analysis & Explicit feedback & - & Error-free channel & History CSI  \\ \hline

Shen \cite{hshca}, control frames' encoding, 2012& Analysis  & - & - & Symmetrical channel & Greedy  \\ \hline

Liao \cite{liao2013uni}, unified up/downlink MAC, 2013& Simulation + Analysis & Implicit feedback & - & Channel reciprocity & FIFO  \\ \hline

Yun \cite{6566748}, multi-APs to multi-STAs, 2013& Testbed & Sequential preambles & Zero Forcing & - & FIFO  \\ \hline

\end{tabular}}
\label{updownmac}
\end{lrbox}
\resizebox{1\textwidth}{!}{\usebox{\tablebox}}
\end{table*}

\subsubsection{Discussions on Integrated Up/down-link MAC}

	\begin{itemize}	
		\item The comments detailed in previous subsections are also valid here. Regarding integrated up/down-link MU-MIMO MAC protocols, we would like to remark that there are still too few works to allow us to categorize them. A first simple classification can be made based on if the AP governs both downlink and uplink communication \cite{hshca}\cite{hlimm}\cite{tkimm}, or the random access is still considered to initiate transmissions from STAs \cite{liao2013uni}\cite{6566748}.
		
		\item Following the IEEE amendment evolution towards 802.11ac, we believe the AP will also play a coordinating role in scheduling uplink MU-MIMO transmissions. 

\item However, there is another major issue that requires further research, which is the need for a joint design of up/down-link MU-MIMO MAC protocols, rather than just simply combining solutions designed independently. The reason is that one-side enhancement could negatively affect the other side. For instance, as pointed out in \cite{liao2013uni}, the optimal parameter for uplink MU-MIMO transmissions does not bring the same benefit to the downlink.
	\end{itemize}

\section{Future Directions}\label{sec:future}
The goal of all the surveyed MAC proposals is to improve MAC efficiency in essence. More specifically, they try to convert the raw data rate brought by the PHY advance (i.e., MU-MIMO) to the MAC throughput as efficiently as possible. 

With this goal in mind, we discuss possible future research directions for MU-MIMO MAC protocols through a micro perspective and a macro perspective, respectively. In the former, we discuss some important performance-affecting aspects that have not been specified in IEEE 802.11ac; we also identify key factors that make the conversion from the PHY data rate to the MAC throughput inefficient, and then give our thoughts on possible ways to improve the MAC efficiency. In the latter, we envisage the MAC's role in facilitating the integration of the future heterogeneous networks.


\subsection{The Micro Perspective: Within MU-MIMO Based WLANs}
\subsubsection{What Does 802.11ac Not Specify?}
The downlink MU-MIMO transmission is one of the most significant features introduced by IEEE 802.11ac. In order to perform multi-user transmissions, the amendment proposes the ECFB scheme to feedback the required CSI, and devises a group identifier (Group-ID) field in the PHY preamble to facilitate grouping STAs. However, the following two important factors are not specified in the amendment.
\begin{itemize}
\item{\textbf{Scope and frequency of CSI feedback}}: \cite{6328529} and \cite{liao2013performance} have shown the significant impact of the CSI feedback on the system performance. A clear dilemma regarding the scope and the frequency of the CSI feedback is that a large scale (e.g., all STAs in the network) or frequent CSI requests will introduce huge overheads, while the opposite way leads to that the rendered channel information might be outdated. Therefore, adaptive algorithms are needed to dynamically adjust the scope and the frequency of CSI feedback.

\item{\textbf{Conditions to group/re-group STAs}}: Although it can be argued that the way of grouping STAs depends on the specific application, a smart grouping algorithm has to be designed to identify those STAs that can be co-scheduled, or on what conditions STAs' re-grouping would be triggered.

\end{itemize}

\subsubsection{How to Improve MAC Efficiency?}
As shown in the paper, significant research efforts have been made to adapt IEEE 802.11 MAC to advances such as the multi-antenna technique. However, the MAC throughput is still much lower than the PHY raw rate (lower than $70\%$ in most cases \cite{ciscoAC}). The throughput loss mainly comes from the so-called overheads, which include the management frames (e.g., association requests/responses), the compulsory idle duration (e.g., the random BO, DIFS/SIFS), control frames (e.g., RTS/CTS/ACK) and frame headers (e.g., PHY preambles, PHY/MAC headers). Other non-overhead factors contributing to the throughput loss include frame collisions and the airtime unfairness caused by low rate STAs that monopolize the channel. These fundamental IEEE 802.11 mechanisms and features limit the MAC efficiency. In a packet capture test conducted in a dense area of Tokyo \cite{akis}, the results show that data frames only account for $23\%$ of all types of frames ($46\%$ management frames, $30\%$ control frames, $1\%$ others); moreover, most of management frames are in the 802.11b format and transmitted at $1$ Mbps to ensure the interoperability.

The newly created High Efficiency WLAN study group-802.11ax is still at its early stage. Here, we give our thoughts on possible solutions to improve the MAC efficiency.
\begin{itemize}
\item{\textbf{Cooperation among multiple bands}}: Management frames, control frames and frames headers are necessary to facilitate correct receptions of data. They can not be eliminated in the current 802.11 communication architecture. At least the evolution from  IEEE 802.11-1997 to 802.11ac, what we have seen is an ever-increasing length of PHY preamble. Cooperation among multiple bands could be an effective way to control overheads. Specifically, a future smart device is likely to be equipped with multiple interfaces operating in multiple bands \cite{singh2011green}. Thus, 802.11ac at $5$ GHz, could be a candidate for the carrier sense and data transmissions across rooms; 802.11ah below $1$ GHz, could be utilized to transmit management and control frames; while 802.11ad at $60$ GHz, could be used for very high speed data transmissions in the line of sight. \cite{lcco} has suggested a possible usage model using 802.11ah for signalling among APs, while 802.11ac for data frames.

\item{\textbf{Revision of the backoff scheme}}: The random BO scheme is a key function employed by IEEE 802.11 to avoid collisions. Unfortunately, collisions can not be eliminated and remain as one of the most degrading factors to the system performance. \cite{jaume2011towards} proposes a collision-free solution, where STAs adopt a deterministic BO instead of a random one after successful transmissions for single-antenna based WLANs. On the other side, works in \cite{li2014cuts} and \cite{sen2011no} suggest to shift the random BO countdown from the time domain to the frequency one by treating OFDM subcarriers as integers. These innovative ideas can be considered to support multi-antenna based WLANs in both downlink and uplink.


\item{\textbf{Uplink MU-MIMO transmissions}}: The Internet traffic has evolved from web browsing and file transfers to a wide variety of applications, which include considerable amount of content-rich files generated by users, such as the video conferencing, social networks and cloud uploading services. Although the enhancement for uplink transmissions has recently gained attention, there is still much space for improvements since the latest amendment-IEEE 802.11ac does not support uplink MU-MIMO transmissions. In addition, there are only a few works that unify MU-MIMO downlink and uplink protocols into a single communication system.

\item{\textbf{Orthogonal FDMA}}:
OFDMA is currently under consideration for next-generation WLANs to improve the system efficiency \cite{gong2014advanced}. The interest on OFDMA is originated from the efficiency loss caused by the 802.11ac channel bonding (up to $160$ MHz). Transmitting over wider channels reduces the transmission duration, though it also magnifies the negative impact of the overheads, since the duration of inter-frame spaces and control packets are constant regardless of the channel width. OFDMA enables WLANs to multiplex transmissions from/to different STAs over a single channel by assigning a different OFDM subcarrier to each transmission. The result is that, transmissions last longer, as they are transmitted at lower rates, but the efficiency of the system is improved, as more data are transmitted in parallel. Regarding MU-MIMO transmissions, using OFDMA would have a positive impact on overheads, since STAs only need to feedback a lower amount of CSI, which is only proportional to the subcarriers of the used subchannel.

\item{\textbf{Full-duplex transmissions}}: The current form of communications in WLANs is half-duplex, namely, transmissions and receptions are allocated to different time slots or frequency bands. The full-duplex transmission has the potential to double the channel capacity by allowing simultaneous transmissions and receptions with the same frequency \cite{aryafar2012midu}\cite{duarte2012design}. In addition, the coming of full-duplex transmissions hints us to rethink the IEEE 802.11 fundamental MAC mechanism-CSMA/CA, which is employed based on the long-held assumptions that (1) wireless devices can not transmit and receive at the same time, and (2) wireless devices can not detect collisions while transmitting \cite{choi2012beyond}. The research on full-duplex transmissions has just started in recent years. Although the challenging part is to cancel self-interference at the full-duplex transceiver, the MAC scheme needs to be revised as well \cite{sahai2011pushing}\cite{zhourctc}.

\item{\textbf{Massive MIMO transmissions}}: Having MU-MIMO already been integrated in the latest standards (e.g., IEEE 802.11ac and 4G LTE), and in order to further reap the benefits of MIMO in a much greater scale, Massive MIMO, with orders of magnitude more antennas equipped at the transmitter, has recently attracted much research attention. Massive MIMO has the promise to increase the spectral efficiency to tens of hundreds of bps/Hz, and simultaneously improve the energy efficiency \cite{larsson2013massive}\cite{lu2013overview}, which enables it to be a very good technology candidate for next-generation wireless standards that target highly-dense user scenarios. The first challenge of employing Massive MIMO is the limited space at the AP, since antennas are required to be placed at the least half wavelength apart. The second challenge is the overhead. IEEE 802.11ac uses ECFB-the explicit way to feedback CSI. The channel sounding time and the CSI volume are proportional to the number of AP's antennas, which would occupy considerable amount of time and bandwidth, especially in the Massive MIMO case. The first challenge is out of the scope of the paper, while the second one needs more research efforts on the precoding, the antenna selection, and exploring the antenna correlation to reduce the channel sounding time and the CSI volume. 

\end{itemize}

\subsection{The Macro Perspective: Integration with Heterogeneous Networks}
It seems to be a trend that in the future there will be a huge and smart network that integrates heterogeneous networks (e.g., WLANs, broadband mobile networks and sensor networks), which means individual network will have to collaborate with others to provide services, rather than just coexist. Obviously, the integration or the inclusion of WLANs require unique changes at the MAC layer. 

\subsubsection{How, who and when to collaborate?}
With the PHY layer's focus on the air interface and the network layer's focus on the routing, the MAC layer plays a role in deciding how to collaborate, who to collaborate and when to collaborate \cite{6189408}\cite{minh2013managing}. For example, imagining a scenario that a sink of a smart metering sensor network requests an AP to forward the collected data to a user's mobile phone, the AP first checks whether the mobile phone is in its vicinity to decide whether to forward the data by itself or to relay the data to other APs. And then, the AP checks the channel condition and the queueing status to decide who and when to transmit the data. Regardless the scale and the type of collaborations, exchanging management frames and control frames are needed to build connections with other entities. Note that today's device-to-device communications already account for a significant part of total wireless traffic, and they are regarded as one of the most important challenges for $5$G networks \cite{hossain2012smart}\cite{hossain2014evolution}. Therefore, the MAC scheme for the integrated network has to take how, who and when to collaborate into account.

\subsubsection{Multi-hop Cooperative MAC}
In addition to the above consideration, it is inevitable for MAC protocols of integrated networks to support multi-hop indirected links. A typical cooperative scenario is that a receiver at the edge of the transmitter's range can benefit from its neighbours' relaying. This opens research questions as follows. First, the selection of optimal relay: (1) who will make the selection, the source or the destination? (2) what is the criteria to choose the optimal relay, signal strength, energy level or security issues? (3) whether the chosen one will agree or be able to relay? Secondly, the timing of relay: (1) immediately after the source transmission, or (2) compete for the channel and then relay. Thirdly, the format of relay: (1) amplify-relay, decode-relay or compress-relay, and (2) single relay or multiple relays  \cite{ju2013survey}. Other issues, such as the hidden-node problem, cooperative diversity, MPT/MPR functionalities and the joint MAC-routing design, also need to be considered \cite{shan2009distributed}\cite{le2008cross}.

\subsubsection{Outdoor and mobile WLANs}
The inclusion of WLANs to outdoors presents some significant challenges to the traditional MAC, as which is designed to support limited mobility. First, due to the user movement and the large scale fading, the outdoor channel is varying faster than that of indoors, even with the same moving speed \cite{HEWPHY}. For that reason, the CSI feedback scheme  needs to be reconsidered to report the channel state timely, while maintaining low channel estimation overheads. Secondly, the cellular operators may offload traffic to public or proprietary WLANs (e.g., the carrier class Wi-Fi \cite{Tataoffload}) in a dense public stadium, in which case, the seamless transfer and the QoS promised by the cellular network need to be assured at the MAC layer. 

\section{Concluding Remarks}\label{sec:conclusions}
The uplink and downlink MU-MIMO MAC protocols for WLANs are investigated and categorized in the paper. Some typical MUD and MUIC techniques for de/pre-coding are sampled, and the requirements for designing MU-MIMO MAC protocols are identified. Based on the study, discussions are carried out to clarify what challenges and future directions could be for designing effective MU-MIMO MAC protocols. 

Despite considerable research has been conducted, there still exists under-explored areas toward simple, yet highly efficient MAC protocols for MU-MIMO based WLANs, especially in the context of the rapid growth of wireless devices. Therefore, we have given some of our thoughts in that regard. We hope this survey paper would help the readers to summarize the current research progress and inspire their future work.

\section*{Acknowledgment}
This work has been supported by the Spanish Government and the Catalan Government under projects TEC2012-32354 (Plan Nacional I+D) and SGR2009\#00617.

The authors are also grateful for the useful technical discussions with Xu Zhang and Shan Zhang at NiuLab of Tsinghua University.

\ifCLASSOPTIONcaptionsoff
  \newpage
\fi



%
\bibliographystyle{ieeetr}	
\bibliography{Survey_refs}

\begin{thebibliography}{100}

\bibitem{ciscoTrend}
Cisco, ``{The Zettabyte Era-Trends and Analysis},'' in {\em Cisco White Paper},
  pp.~1--19, 2013.

\bibitem{IEEEac}
``{IEEE Draft Standard for Information Technology--LAN/MAN--Part 11: Wireless
  LAN Medium Access Control and Physical Layer Specifications--Amendment:
  Enhancements for Very High Throughput for Operation in Bands Below 6GHz},''
  {\em IEEE P802.11ac/D6.0}, pp.~1--446, 2013.

\bibitem{goldsmith2003capacity}
A.~Goldsmith, S.~A. Jafar, N.~Jindal, and S.~Vishwanath, ``{Capacity limits of
  MIMO channels},'' {\em IEEE Journal on Selected Areas in Communications},
  vol.~21, no.~5, pp.~684--702, 2003.

\bibitem{gcaicl}
G.~Caire and S.~Shamai, ``{On the achievable throughput of a multiantenna
  Gaussian broadcast channel},'' {\em IEEE Transactions on Information Theory},
  vol.~49, no.~7, pp.~1691--1706, 2003.

\bibitem{dgess}
D.~Gesbert, M.~Kountouris, R.~W. Heath, C.-B. Chae, and T.~Salzer, ``{Shifting
  the MIMO paradigm: From Single User to Multiuser Communications},'' {\em IEEE
  Signal Processing Magazine}, vol.~24, no.~5, pp.~36--46, 2007.

\bibitem{spencer2004zero}
Q.~H. Spencer, A.~L. Swindlehurst, and M.~Haardt, ``Zero-forcing methods for
  downlink spatial multiplexing in multiuser mimo channels,'' {\em IEEE
  Transactions on Signal Processing}, vol.~52, no.~2, pp.~461--471, 2004.

\bibitem{cbpav}
C.~B. Peel, B.~M. Hochwald, and A.~L. Swindlehurst, ``A vector-perturbation
  technique for near-capacity multiantenna multiuser communication-part i:
  channel inversion and regularization,'' {\em IEEE Transactions on
  Communications}, vol.~53, no.~1, pp.~195--202, 2005.

\bibitem{jmie}
J.~Mietzner, R.~Schober, L.~Lampe, W.~H. Gerstacker, and P.~A. Hoeher,
  ``Multiple-antenna techniques for wireless communications-a comprehensive
  literature survey,'' {\em IEEE Communications Surveys \& Tutorials}, vol.~11,
  no.~2, pp.~87--105, 2009.

\bibitem{abramson1970aloha}
N.~Abramson, ``{THE ALOHA SYSTEM: another alternative for computer
  communications},'' in {\em Proceedings of the AFIPS Fall Joint Computer
  Conference}, pp.~281--285, ACM, 1970.

\bibitem{ieesast}
IEEE, ``{IEEE STANDARDS BOARD OPERATIONS MANUAL}.''
  \url{http://standards.ieee.org/develop/policies/opman/sect1.html}.
\newblock [Accessed 15 October 2014].

\bibitem{IEEE12}
``{IEEE Standard for Information technology--Telecommunications and information
  exchange between systems Local and metropolitan area networks--Specific
  requirements Part 11: Wireless LAN Medium Access Control (MAC) and Physical
  Layer (PHY) Specifications},'' {\em IEEE Std 802.11-2012}, pp.~1--2793, 2012.

\bibitem{IEEE09n}
``{IEEE Standard for Information technology--LAN/MAN-- Specific requirements--
  Part 11: Wireless LAN Medium Access Control and Physical Layer Specifications
  Amendment 5: Enhancements for Higher Throughput},'' {\em IEEE 802.11n},
  pp.~1--565, 2009.

\bibitem{11ax}
{IEEE 802.11ax}, ``{Project IEEE 802.11ax High Efficiency WLAN (HEW)}.''
  \url{http://grouper.ieee.org/groups/802/11/Reports/tgax_update.htm}.
\newblock [Accessed 15 October 2014].

\bibitem{lcum}
L.~Cariou, ``{Usage models for IEEE 802.11 High Efficiency WLAN study group
  (HEW SG)}.''
  \url{https://mentor.ieee.org/802.11/dcn/13/11-13-0657-06-0hew-hew-sg-usage-models-and-requirements-liaison-with-wfa.ppt}.
\newblock [Accessed 15 October 2014].

\bibitem{gong2014advanced}
M.~X. Gong, B.~Hart, and S.~Mao, ``{Advanced Wireless LAN Technologies: IEEE
  802.11 ac and Beyond},'' {\em ACM Mobile Computing and Communications Review
  (MC2R)}, 2014.

\bibitem{lwpax}
{L. Wang, H. Lou, H. Zhang, Y. Sun, L. Chu, Z. Lan, J. Zhang, H. Kang, S. Chang
  and R. Taori}, ``{Proposed 802.11ax Functional Requirements}.''
  \url{https://mentor.ieee.org/802.11/dcn/14/11-14-0567-03-00ax-proposed-tgax-functional-requirements.doc}.
\newblock [Accessed 15 October 2014].

\bibitem{11ad}
{IEEE 802.11ad}, ``{Very High Throughput in 60 GHz}.''
  \url{http://www.ieee802.org/11/Reports/tgad_update.htm}.
\newblock [Accessed 15 October 2014].

\bibitem{11ah}
{IEEE 802.11 11ah}, ``{Proposed TGah Draft Amendment}.''
  \url{http://www.ieee802.org/11/Reports/tgah_update.htm}.
\newblock [Accessed 15 October 2014].

\bibitem{adame2013capacity}
T.~Adame, A.~Bel, B.~Bellalta, J.~Barcelo, J.~Gonzalez, and M.~Oliver,
  ``{Capacity Analysis of IEEE 802.11 ah WLANs for M2M Communications},'' in
  {\em Multiple Access Communcations}, pp.~139--155, Springer, 2013.

\bibitem{adame2014ieee}
T.~Adame, A.~Bel, B.~Bellalta, J.~Barcelo, and M.~Oliver, ``Ieee 802.11 ah: The
  wi-fi approach for m2m communications,'' {\em IEEE Wireless Communications},
  2014.

\bibitem{IEEE05e}
``{IEEE Standard for Information Technology--LAN/MAN--Part 11: Wireless LAN
  Medium Access Control and Physical Layer Specifications--Amendment: Medium
  access control (MAC) Enhancements for Quality of Service},'' {\em IEEE
  802.11e}, pp.~1--211, 2005.

\bibitem{gbu8}
G.~Bianchi, I.~Tinnirello, and L.~Scalia, ``{Understanding 802.11 e
  contention-based prioritization mechanisms and their coexistence with legacy
  802.11 stations},'' {\em Network, IEEE}, vol.~19, no.~4, pp.~28--34, 2005.

\bibitem{charfi2013phy}
E.~Charfi, L.~Chaari, and L.~Kamoun, ``{PHY/MAC enhancements and QoS mechanisms
  for very high throughput WLANs: a survey},'' {\em IEEE Communications Surveys
  \& Tutorials}, 2013.

\bibitem{jiang2007multiuser}
M.~Jiang and L.~Hanzo, ``{Multiuser MIMO-OFDM for next-generation wireless
  systems},'' {\em Proceedings of the IEEE}, vol.~95, no.~7, pp.~1430--1469,
  2007.

\bibitem{khalid2010advances}
F.~Khalid and J.~Speidel, ``{Advances in MIMO Techniques for Mobile
  Communications-A Survey},'' {\em Int'l J. of Communications, Network and
  System Sciences}, vol.~3, no.~3, pp.~213--252, 2010.

\bibitem{paulraj2003introduction}
A.~Paulraj, R.~Nabar, and D.~Gore, {\em {Introduction to space time wireless
  communications}}.
\newblock Cambridge university press, 2003.

\bibitem{costa1983writing}
M.~Costa, ``{Writing on dirty paper (Correspondence)},'' {\em IEEE Transactions
  on Information Theory}, vol.~29, no.~3, pp.~439--441, 1983.

\bibitem{dlova}
D.~J. Love, R.~W. Heath, V.~K. Lau, D.~Gesbert, B.~D. Rao, and M.~Andrews, ``An
  overview of limited feedback in wireless communication systems,'' {\em IEEE
  Journal on Selected Areas in Communications}, vol.~26, no.~8, pp.~1341--1365,
  2008.

\bibitem{lou2013comparison}
H.~Lou, M.~Ghosh, P.~Xia, and R.~Olesen, ``{A comparison of implicit and
  explicit channel feedback methods for MU-MIMO WLAN systems},'' in {\em
  PIMRC}, pp.~419--424, IEEE, 2013.

\bibitem{gong2010training}
M.~X. Gong, E.~Perahia, R.~Want, and S.~Mao, ``{Training protocols for
  multi-user MIMO wireless LANs},'' in {\em PIMRC}, pp.~1218--1223, IEEE, 2010.

\bibitem{jhup}
H.~Jin, B.~C. Jung, H.~Y. Hwang, and D.~K. Sung, ``{Performance Comparison of
  Uplink WLANs with Single-User and Multi-User MIMO Schemes},'' in {\em WCNC},
  pp.~1854--1859, 2008.

\bibitem{BabichC10}
F.~Babich and M.~Comisso, ``{Theoretical Analysis of Asynchronous Multi-packet
  Reception in 802.11 Networks},'' {\em IEEE Transactions on Communications},
  vol.~58, no.~6, pp.~1782--1794, 2010.

\bibitem{PZheng}
P.~X. Zheng, Y.~J. Zhang, and S.~C. Liew, ``{Multipacket Reception in Wireless
  Local Area Networks},'' in {\em ICC}, vol.~8, pp.~3670--3675, 2006.

\bibitem{rldde}
R.~Liao, B.~Bellalta, C.~Cano, and M.~Oliver, ``{DCF/DSDMA: Enhanced DCF with
  SDMA Downlink Transmissions for WLANs},'' in {\em BCFIC}, pp.~96--102, 2011.

\bibitem{bellalta2012performance}
B.~Bellalta, J.~Barcelo, D.~Staehle, A.~Vinel, and M.~Oliver, ``{On the
  performance of packet aggregation in IEEE 802.11 ac MU-MIMO WLANs},'' {\em
  IEEE Communications Letters}, vol.~16, no.~10, pp.~1588--1591, 2012.

\bibitem{liao2013uni}
R.~Liao, B.~Bellalta, T.~C. Minh, J.~Barcelo, and M.~Oliver, ``{Uni-MUMAC: A
  Unified Down/Up-link MU-MIMO MAC Protocol for IEEE 802.11 ac WLANs},'' {\em
  arXiv preprint arXiv:1309.5049}, 2013.

\bibitem{6566748}
S.~Yun, L.~Qiu, and A.~Bhartia, ``{Multi-point to multi-point MIMO in wireless
  LANs},'' in {\em INFOCOM}, pp.~125--129, 2013.

\bibitem{LCai}
L.~X. Cai, H.~Shan, W.~Zhuang, X.~Shen, J.~W. Mark, and Z.~Wang, ``{A
  Distributed Multi-User MIMO MAC Protocol for Wireless Local Area Networks},''
  in {\em GLOBECOM}, pp.~4976--4980, 2008.

\bibitem{mxgac}
M.~X. Gong, E.~Perahia, R.~Stacey, R.~Want, and S.~Mao, ``{A CSMA/CA MAC
  Protocol for Multi-User MIMO Wireless LANs},'' in {\em GLOBECOM}, pp.~1--6,
  2010.

\bibitem{zhang2010employing}
Z.~Zhang, S.~Bronson, J.~Xie, and H.~Wei, ``{Employing the
  one-sender-multiple-receiver technique in wireless LANs},'' in {\em INFOCOM},
  pp.~1--9, IEEE, 2010.

\bibitem{hshca}
H.~Shen, S.~Lv, Y.~Sun, X.~Dong, X.~Wang, and X.~Zhou, ``{Concurrent Access
  Control Using Subcarrier Signature in Heterogeneous MIMO-Based WLAN},'' in
  {\em MACOM}, pp.~109--121, 2012.

\bibitem{whuaj}
W.~L. Huang, K.~Ben~Letaief, and Y.~J. Zhang, ``{Joint Channel State Based
  Random Access and Adaptive Modulation in Wireless LANs with Multi-Packet
  Reception},'' {\em IEEE Transactions on Wireless Communications}, vol.~7,
  no.~11, pp.~4185--4197, 2008.

\bibitem{tatae}
T.~Tandai, H.~Mori, K.~Toshimitsu, and T.~Kobayashi, ``{An efficient uplink
  multiuser MIMO protocol in IEEE 802.11 WLANs},'' in {\em PIMRC},
  pp.~1153--1157, IEEE, 2009.

\bibitem{EKar}
E.~Kartsakli, N.~Zorba, L.~Alonso, and C.~V. Verikoukis, ``{Multiuser MAC
  Protocols for 802.11n Wireless Networks},'' in {\em ICC}, pp.~1--5, 2009.

\bibitem{zhao2009applying}
M.~Zhao, M.~Ma, and Y.~Yang, ``{Applying opportunistic medium access and
  multiuser MIMO techniques in multi-channel multi-radio WLANs},'' {\em Mobile
  Networks and Applications}, vol.~14, no.~4, pp.~486--507, 2009.

\bibitem{hlimm}
H.~Li, A.~Attar, and V.~C.~M. Leung, ``{Multi-User Medium Access Control in
  Wireless Local Area Network},'' in {\em WCNC}, pp.~1--6, 2010.

\bibitem{CAnt}
C.~Anton-Haro, P.~Svedman, M.~Bengtsson, A.~Alexiou, and A.~Gameiro,
  ``{Cross-layer scheduling for multi-user MIMO systems},'' {\em IEEE
  Communications Magazine}, vol.~44, no.~9, pp.~39--45, 2006.

\bibitem{vkawa}
V.~Kawadia and P.~Kumar, ``A cautionary perspective on cross-layer design,''
  {\em IEEE Wireless Communications}, vol.~12, no.~1, pp.~3--11, 2005.

\bibitem{whuac}
W.~L. Huang, K.~Letaief, and Y.~J. Zhang, ``{Cross-layer multi-packet reception
  based medium access control and resource allocation for space-time coded
  MIMO/OFDM},'' {\em IEEE Transactions on Wireless Communications}, vol.~7,
  no.~9, pp.~3372--3384, 2008.

\bibitem{yu2013resource}
X.~Yu, P.~Navaratnam, and K.~Moessner, ``{Resource Reservation Schemes for IEEE
  802.11-Based Wireless Networks: A Survey},'' {\em IEEE Communications Surveys
  \& Tutorials}, vol.~15, no.~3, pp.~1042--1061, 2013.

\bibitem{840210}
G.~Bianchi, ``{Performance analysis of the IEEE 802.11 distributed coordination
  function},'' {\em IEEE Journal on Selected Areas in Communications}, vol.~18,
  no.~3, pp.~535--547, 2000.

\bibitem{kutse}
K.~Tan, H.~Liu, J.~Fang, W.~Wang, J.~Zhang, M.~Chen, and G.~M. Voelker, ``{SAM:
  enabling practical spatial multiple access in wireless LAN},'' in {\em
  INFOCOM}, pp.~49--60, ACM, 2009.

\bibitem{sora}
{Microsoft}, ``{Research Software Radio (SORA)}.''
  \url{http://research.microsoft.com/en-us/projects/sora}.
\newblock [Accessed 15 October 2014].

\bibitem{ettefagh2011performance}
A.~Ettefagh, M.~Kuhn, C.~E{\c{s}}li, and A.~Wittneben, ``{Performance analysis
  of distributed cluster-based MAC protocol for multiuser MIMO wireless
  networks},'' {\em EURASIP Journal on Wireless Communications and Networking},
  vol.~2011, no.~1, pp.~1--14, 2011.

\bibitem{mukhopadhyay2012acknowledgement}
A.~Mukhopadhyay, N.~B. Mehta, and V.~Srinivasan, ``{Acknowledgement-aware MPR
  MAC protocol for distributed WLANs: Design and analysis},'' in {\em
  GLOBECOM}, pp.~5087--5092, IEEE, 2012.

\bibitem{6566751}
T.-H. Lin and H.~Kung, ``Concurrent channel access and estimation for scalable
  multiuser mimo networking,'' in {\em INFOCOM}, pp.~140--144, April 2013.

\bibitem{candes2008introduction}
E.~J. Cand{\`e}s and M.~B. Wakin, ``An introduction to compressive sampling,''
  {\em IEEE Signal Processing Magazine}, vol.~25, no.~2, pp.~21--30, 2008.

\bibitem{6807572}
S.~Wu, W.~Mao, and X.~Wang, ``{Performance Study on a CSMA/CA-Based MAC
  Protocol for Multi-User MIMO Wireless LANs},'' {\em IEEE Transactions on
  Wireless Communications}, vol.~13, no.~6, pp.~3153--3166, 2014.

\bibitem{kuo2014leader}
T.-W. Kuo, K.-C. Lee, K.-J. Lin, and M.-J. Tsai, ``{Leader-Contention-Based
  User Matching for 802.11 Multiuser MIMO Networks},'' {\em IEEE Transactions
  on Wireless Communications 13(8): 4389-4400}.

\bibitem{chong2013introduction}
E.~K. Chong and S.~H. Zak, {\em An introduction to optimization}, vol.~76.
\newblock John Wiley \& Sons, 2013.

\bibitem{li2014cuts}
H.~Li, K.~Wu, Q.~Zhang, and L.~M. Ni, ``{CUTS: Improving channel utilization in
  both time and spatial domains in WLANs},'' {\em IEEE Transactions on Parallel
  and Distributed Systems, Vol. 25, No. 6}, 2014.

\bibitem{barghi2011mimo}
S.~Barghi, H.~Jafarkhani, and H.~Yousefi'zadeh, ``{MIMO-assisted MPR-aware MAC
  design for asynchronous WLANs},'' {\em IEEE/ACM Transactions on Networking},
  vol.~19, no.~6, pp.~1652--1665, 2011.

\bibitem{SZhou}
S.~Zhou and Z.~Niu, ``{Distributed Medium Access Control with SDMA Support for
  WLANs},'' {\em IEICE Transactions}, vol.~93-B, no.~4, pp.~961--970, 2010.

\bibitem{rlidu}
R.~Liao, B.~Bellalta, and M.~Oliver, ``{DCF/USDMA: Enhanced DCF for Uplink SDMA
  Transmissions in WLANs},'' in {\em IWCMC}, pp.~263--268, 2012.

\bibitem{yjzmr}
Y.~J. Zhang, ``{Multi-round contention in wireless LANs with multipacket
  reception},'' {\em IEEE Transactions on Wireless Communications}, vol.~9,
  pp.~1503--1513, Apr. 2010.

\bibitem{chow1971great}
Y.~S. Chow, H.~Robbins, and D.~Siegmund, {\em Great expectations: The theory of
  optimal stopping}.
\newblock Houghton Mifflin Boston, 1971.

\bibitem{6302115}
D.~Jung, R.~Kim, and H.~Lim, ``{Asynchronous Medium Access Protocol for
  Multi-User MIMO Based Uplink WLANs},'' {\em IEEE Transactions on
  Communications}, vol.~60, no.~12, pp.~3745--3754, 2012.

\bibitem{tnup}
{T. Nguyen, L. Lanante, H. Ochi, T. Uwai and Y. Nagao}, ``{Uplink multi-user
  MAC protocol for 11ax}.''
  \url{https://mentor.ieee.org/802.11/dcn/14/11-14-0598-00-00ax-uplink-multi-user-mac-protocol-for-11ax.pptx}.
\newblock [Accessed 15 October 2014].

\bibitem{zlup}
{Z. Lan, Y. Li, D. Yang, J. Zhang, R. Luo, P. Loc and P. Barber}, ``{Frame
  Exchange Control for Uplink Multi-user transmission}.''
  \url{https://mentor.ieee.org/802.11/dcn/14/11-14-1190-03-00ax-frame-exchange-control-for-uplink-multi-user-transmission.pptx}.
\newblock [Accessed 15 October 2014].

\bibitem{cai2006voice}
L.~X. Cai, X.~Shen, J.~W. Mark, L.~Cai, and Y.~Xiao, ``Voice capacity analysis
  of wlan with unbalanced traffic,'' {\em Vehicular Technology, IEEE
  Transactions on}, vol.~55, no.~3, pp.~752--761, 2006.

\bibitem{usrp}
{Ettus}, ``{Universal Software Radio Peripheral (USRP)}.''
  \url{http://www.ettus.com}.
\newblock [Accessed 15 October 2014].

\bibitem{bellalta2009space}
B.~Bellalta and M.~Oliver, ``{A space-time batch-service queueing model for
  multi-user MIMO communication systems},'' in {\em ACM international
  conference on Modeling, analysis and simulation of wireless and mobile
  systems}, pp.~357--364, ACM, 2009.

\bibitem{redieteab2012mu}
G.~Redieteab, L.~Cariou, P.~Christin, and J.-F. H{\'e}lard, ``{SU/MU-MIMO in
  IEEE 802.11 ac: PHY+ MAC performance comparison for single antenna
  stations},'' in {\em Wireless Telecommunications Symposium}, pp.~1--5, IEEE,
  2012.

\bibitem{11acChannel}
{G. Breit, et al.}, ``{IEEE P802.11 Wireless LANs TGac Channel Model
  Addendum}.''
  \url{https://mentor.ieee.org/802.11/dcn/09/11-09-0308-03-00ac-tgac-channel-model-addendum-document.doc}.
\newblock [Accessed 15 October 2014].

\bibitem{jcpc}
J.~Cha, H.~Jin, B.~C. Jung, and D.~K. Sung, ``{Performance Comparison of
  Downlink User Multiplexing Schemes in IEEE 802.11ac: Multi-user MIMO vs.
  Frame Aggregation},'' in {\em WCNC}, pp.~1514--1519, 2012.

\bibitem{hbaem}
H.~V. Balan, R.~Rogalin, A.~Michaloliakos, K.~Psounis, and G.~Caire,
  ``Achieving high data rates in a distributed mimo system,'' in {\em MOBICOM},
  pp.~41--52, ACM, 2012.

\bibitem{warp}
{Rice University}, ``{Wireless Open-Access Research Platform (WARP)}.''
  \url{http://warp.rice.edu/trac/wiki/about}.
\newblock [Accessed 15 October 2014].

\bibitem{czme}
C.~Zhu, A.~Bhatt, Y.~Kim, O.~Aboul-magd, and C.~Ngo, ``{MAC Enhancements for
  Downlink Multi-user MIMO Transmission in Next Generation WLAN},'' in {\em
  CCNC}, pp.~832--837, 2012.

\bibitem{ji2014cooperative}
B.~Ji, K.~Song, Y.~Hu, and H.~Chen, ``{Cooperative Transmission Mechanisms in
  Next Generation WiFi: IEEE 802.11 ac},'' {\em International Journal of
  Distributed Sensor Networks}, vol.~2014, 2014.

\bibitem{6328529}
G.~Redieteab, L.~Cariou, P.~Christin, and J.~F. Helard, ``{PHY+MAC channel
  sounding interval analysis for IEEE 802.11ac MU-MIMO},'' in {\em ISWCS},
  pp.~1054--1058, 2012.

\bibitem{liao2013performance}
R.~Liao, B.~Bellalta, J.~Barcelo, V.~Valls, and M.~Oliver, ``{Performance
  analysis of IEEE 802.11 ac wireless backhaul networks in saturated
  conditions},'' {\em EURASIP Journal on Wireless Communications and
  Networking}, vol.~2013, no.~1, pp.~1--14, 2013.

\bibitem{maraslis2012ieee}
K.~Maraslis, P.~Chatzimisios, and A.~Boucouvalas, ``{IEEE 802.11 aa:
  Improvements on video transmission over Wireless LANs},'' in {\em ICC},
  pp.~115--119, IEEE, 2012.

\bibitem{tkimm}
T.~Kim and N.~Vaidya, ``{MAC Protocol Design for Multiuser MIMO Wireless
  Networks},'' {Technical Report}, University of Illinois at Urbana-Champaign,
  2008.

\bibitem{ciscoAC}
Cisco, ``{802.11ac: The Fifth Generation of Wi-Fi},'' in {\em Cisco White
  Paper}, pp.~1--25, 2012.

\bibitem{akis}
{A. Kishida, M. Iwabuchi, Y. Inoue, Y. Asai, Y. Takatori, T. Shintaku, T.
  Sakata and A. Yamada}, ``{Issues of Low-Rate Transmission}.''
  \url{https://mentor.ieee.org/802.11/dcn/13/11-13-0801-01-0hew-issues-of-low-rate-transmission.pptx}.
\newblock [Accessed 15 October 2014].

\bibitem{singh2011green}
H.~Singh, J.~Hsu, L.~Verma, S.~S. Lee, and C.~Ngo, ``{Green operation of
  multi-band wireless LAN in 60 GHz and 2.4/5 GHz},'' in {\em CCNC},
  pp.~787--792, IEEE, 2011.

\bibitem{lcco}
T.~D. Laurent~Cariou and J.-P.~L. Rouzic, ``{Carrier-oriented WIFI for cellular
  offload}.''
  \url{https://mentor.ieee.org/802.11/dcn/12/11-12-0910-00-0wng-carrier-oriented-wifi-cellular-offload.ppt}.
\newblock [Accessed 15 October 2014].

\bibitem{jaume2011towards}
J.~Barcelo, B.~Bellalta, C.~Cano, A.~Sfairopoulou, M.~Oliver, and K.~Verma,
  ``{Towards a collision-free WLAN: dynamic parameter adjustment in
  CSMA/E2CA},'' {\em EURASIP Journal on Wireless Communications and
  Networking}, vol.~2011, pp.~1--11, 2011.

\bibitem{sen2011no}
S.~Sen, R.~Roy~Choudhury, and S.~Nelakuditi, ``No time to countdown: Migrating
  backoff to the frequency domain,'' in {\em Proceedings of the 17th annual
  international conference on Mobile computing and networking}, pp.~241--252,
  ACM, 2011.

\bibitem{aryafar2012midu}
E.~Aryafar, M.~A. Khojastepour, K.~Sundaresan, S.~Rangarajan, and M.~Chiang,
  ``{MIDU: enabling MIMO full duplex},'' in {\em MobiCom}, pp.~257--268, ACM,
  2012.

\bibitem{duarte2012design}
M.~Duarte, A.~Sabharwal, V.~Aggarwal, R.~Jana, K.~Ramakrishnan, C.~Rice, and
  N.~Shankaranarayanan, ``{Design and Characterization of a Full-Duplex
  Multiantenna System for WiFi Networks},'' {\em IEEE Transactions on Vehicular
  Technology}, vol.~63, no.~3, pp.~1160--1177, 2014.

\bibitem{choi2012beyond}
J.~I. Choi, S.~Hong, M.~Jain, S.~Katti, P.~Levis, and J.~Mehlman, ``{Beyond
  full duplex wireless},'' in {\em ASILOMAR}, pp.~40--44, IEEE, 2012.

\bibitem{sahai2011pushing}
A.~Sahai, G.~Patel, and A.~Sabharwal, ``{Pushing the limits of full-duplex:
  Design and real-time implementation},'' {\em arXiv preprint arXiv:1107.0607},
  2011.

\bibitem{zhourctc}
W.~Zhou, K.~Srinivasan, and P.~Sinha, ``{RCTC: Rapid concurrent transmission
  coordination in full DuplexWireless networks},'' in {\em ICNP}, pp.~1--10,
  2013.

\bibitem{larsson2013massive}
E.~G. Larsson, O.~Edfors, F.~Tufvesson, and T.~L. Marzetta, ``Massive mimo for
  next generation wireless systems,'' {\em arXiv preprint arXiv:1304.6690},
  2013.

\bibitem{lu2013overview}
L.~Lu, G.~Li, A.~Swindlehurst, A.~Ashikhmin, and R.~Zhang, ``An overview of
  massive mimo: Benefits and challenges,'' 2013.

\bibitem{6189408}
W.~Zhuang and M.~ISMAIL, ``{Cooperation in wireless communication networks},''
  {\em IEEE Wireless Communications}, vol.~19, no.~2, pp.~10--20, 2012.

\bibitem{minh2013managing}
T.~C. Minh, B.~Bellalta, S.~Oechsner, R.~Liao, and M.~Oliver, ``{Managing
  Heterogeneous WSNs in Smart Cities: Challenges and Requirements},'' {\em
  arXiv preprint arXiv:1310.6901}, 2013.

\bibitem{hossain2012smart}
E.~Hossain, Z.~Han, and H.~V. Poor, {\em {Smart grid communications and
  networking}}.
\newblock Cambridge University Press, 2012.

\bibitem{hossain2014evolution}
E.~Hossain, M.~Rasti, H.~Tabassum, and A.~Abdelnasser, ``Evolution toward 5g
  multi-tier cellular wireless networks: An interference management
  perspective,'' {\em Wireless Communications, IEEE}, vol.~21, pp.~118--127,
  June 2014.

\bibitem{ju2013survey}
P.~Ju, W.~Song, and D.~Zhou, ``Survey on cooperative medium access control
  protocols,'' {\em IET Communications}, vol.~7, no.~9, pp.~893--902, 2013.

\bibitem{shan2009distributed}
H.~Shan, W.~Zhuang, and Z.~Wang, ``Distributed cooperative mac for multihop
  wireless networks,'' {\em IEEE Communications Magazine}, vol.~47, no.~2,
  pp.~126--133, 2009.

\bibitem{le2008cross}
L.~Le and E.~Hossain, ``{Cross-layer optimization frameworks for multihop
  wireless networks using cooperative diversity},'' {\em IEEE Transactions on
  Wireless Communications}, vol.~7, no.~7, pp.~2592--2602, 2008.

\bibitem{HEWPHY}
{W. Lee, et al.}, ``{HEW SG PHY Considerations For Outdoor Environment}.''
  \url{https://mentor.ieee.org/802.11/dcn/13/11-13-0536-00-0hew-hew-sg-phy-considerations-for-outdoor-environment.pptx}.
\newblock [Accessed 15 October 2014].

\bibitem{Tataoffload}
R.~Agarwal and A.~Tomer, ``{Carrier Wi-Fi Offload: Charting the Road Ahead},''
  in {\em Tata Consultancy Services White Paper}, pp.~1--14, 2013.

\end{thebibliography}

%

\end{document}